\documentclass[11pt]{article}
\pdfoutput=1 
\usepackage{setspace}
\usepackage{fancyhdr}
\usepackage{lastpage}
\usepackage{extramarks}
\usepackage{chngpage}
\usepackage{soul}
\usepackage{graphicx}
\usepackage{float,wrapfig}
\usepackage{ifthen}
\usepackage{listings}
\usepackage{courier}
\usepackage{xcolor}
\usepackage{arxiv}
\usepackage{booktabs}
\usepackage{hyperref}
\hypersetup{hidelinks}

\usepackage{url}
\usepackage{pdfpages} %Multi PDF pages

\usepackage{caption}
\usepackage{subcaption}

\usepackage{amsmath,amsfonts,amsthm,amssymb}
\usepackage[utf8]{inputenc}
\usepackage[english]{babel}
\usepackage{bbm} %bold numbers -> Indicator function

\usepackage{mathtools}
\mathtoolsset{showonlyrefs=true} %Show only numbering for equations that are referenced

\usepackage[linesnumbered,vlined, ruled]{algorithm2e}

\usepackage{array}

\newcolumntype{C}[1]{>{\centering\arraybackslash}p{#1}}

\usepackage[acronym]{glossaries}
\makeglossaries

\usepackage{nomencl}
\makenomenclature

\usepackage{natbib}
\graphicspath{{_figs/}}

\usepackage{comment}

\newtheorem{definition}{Definition}[section]

\newcommand{\beq}{\begin{equation}}
\newcommand{\eeq}{\end{equation}}
\newcommand{\bea}{\begin{eqnarray}}
\newcommand{\eea}{\end{eqnarray}}
\newcommand{\ba}{\begin{array}}
\newcommand{\ea}{\end{array}}
\newcommand{\bi}{\begin{itemize}}
\newcommand{\ei}{\end{itemize}}
\newcommand{\ben}{\begin{enumerate}}
\newcommand{\een}{\end{enumerate}}

\definecolor{dkgreen}{rgb}{0,0.6,0}
\definecolor{gray}{rgb}{0.5,0.5,0.5}
\definecolor{mauve}{rgb}{0.58,0,0.82}

\title{Sequential Bayesian Learning for Hidden Semi-Markov Models}

\author{
 Patrick Aschermayr, Konstantinos Kalogeropoulos\\
 Department of Statistics\\
  LSE\\
  \texttt{p.aschermayr@lse.ac.uk}; \texttt{k.kalogeropoulos@lse.ac.uk}  \\
}
\begin{document}
\maketitle

%%%%%%%%%%%%%%%%%%%%%%%%%%%
% Set glossary and Nomenclature
%%%%%%%%%%%%%%%%%%%%%%%%%%%% Nomenclature
% makeindex Homework_Template.nlo -s nomencl.ist -o Homework_Template.nls
% \nomenclature[1]{$HMM$}{Hidden Markov model}
% \nomenclature[1]{$MCMC$}{Markov chain Monte Carlo}
% \nomenclature[1]{$MH$}{Metropolis Hastings}
% \nomenclature[1]{$EM$}{Expectation maximization}
% \nomenclature[1]{$PMMH$}{Particle marginal Metropolis Hastings}
% \nomenclature[1]{$PF$}{Particle Filter}
% \nomenclature[1]{$PG$}{Particle Gibbs}
% \nomenclature[1]{$SIR$}{Sequential importance resampling}
% \nomenclature[1]{$SMC$}{Sequential Monte Carlo}

%  \newglossaryentry{formula}
%  {
%          name=formula,
%          description={A mathematical expression}
%  }
\newacronym{ssm}{SSM}{State Space Model}

\newacronym{mc}{MC}{Monte Carlo}
\newacronym{ibis}{IBIS}{Iterated Batch Importance Sampling}
\newacronym{smc}{SMC}{Sequential Monte Carlo}
\newacronym{smc2}{SMC2}{Sequential Monte Carlo Squared}
%\newacronym{gibbs}{Gibbs}{Gibbs}

\newacronym{mcmc}{MCMC}{Markov Chain Monte Carlo}
\newacronym{hmc}{HMC}{Hamiltonian Monte Carlo}
\newacronym{nuts}{NUTS}{No U-Turn Sampling}
\newacronym{pmh}{PMH}{Particle Metropolis Hastings}
\newacronym{pgibbs}{PGIBBS}{Particle Gibbs}
\newacronym{pgas}{PGAS}{Particle Gibbs with ancestors sampling}
\newacronym{pmcmc}{PMCMC}{Particle MCMC}

\newacronym{pf}{PF}{Particle Filter}
\newacronym{cpf}{CPF}{Conditional Particle Filter}
\newacronym{bpf}{BPF}{Bootstrap Particle Filter}
\newacronym{apf}{APF}{Auxiliary Particle Filter}

\newacronym{hmm}{HMM}{Hidden Markov Model}
\newacronym{hsmm}{HSMM}{Hidden Semi-Markov Model}
\newacronym{edhmm}{EDHMM}{Explicit-duration Hidden Markov Model}
\newacronym{arhsmm}{ARHSMM}{Autoregressive Hidden Semi-Markov Model}
\newacronym{sv}{SV}{Stochastic Volatility}

\newacronym{hsmm-egarch}{HSMM-eGARCH}{Hidden Semi-Markov Model with eGarch volatility dynamics}

\newacronym{svm}{SVM}{Stochastic Volatility Model}
\newacronym{mjm}{MJM}{Markov Jump Model}
\newacronym{garch}{GARCH}{Generalized Autoregressive Conditional Heteroskedasticity}
\newacronym{egarch}{eGARCH}{Exponential Generalized Autoregressive Conditional Heteroskedasticity}

\newacronym{crps}{CRPS}{Continuous Ranked Probability Score}
\newacronym{bmi}{BMI}{Basic Marginal Likelihood Identity}
\newacronym{is}{IS}{Importance Sampling}
\newacronym{sis}{SIS}{Sequential Importance Sampling}
\newacronym{sir}{SIR}{Sequential Importance Resampling}
\newacronym{ci}{CI}{Credible Interval}
\newacronym{pl}{PL}{Predictive Likelihood}
\newacronym{bf}{BF}{Bayes Factor}
\newacronym{clpbf}{CLPBF}{Cumulative Log Predictive Bayes Factor}
\newacronym{ar}{AR}{Autoregressive}
\newacronym{ess}{ESS}{Effective Sample Size}
\newacronym{pacf}{PACF}{Partial Autocorrelation Function}
\glsaddall

%%%%%%%%%%%%%%%%%%%%%%%%%%%
% Set Abstract
\begin{abstract}
In this paper, we explore the class of the \acrfull{hsmm}, a flexible extension of the popular \acrfull{hmm} that allows the underlying stochastic process to be a semi-Markov chain. \acrshort{hsmm}s are typically used less frequently than their basic \acrshort{hmm} counterpart due to the increased computational challenges when evaluating the likelihood function. 
Moreover, while both models are sequential in nature, parameter estimation is mainly conducted via batch estimation methods. 
Thus, a major motivation of this paper is to provide methods to estimate \acrshort{hsmm}s (1) in a computationally feasible time,
(2) in an exact manner, i.e. only subject to Monte Carlo error,
and (3) in a sequential setting. 
We provide and verify an efficient computational scheme for Bayesian parameter estimation on \acrshort{hsmm}s. Additionally, we explore the performance of \acrshort{hsmm}s 
on the VIX time series using \acrfull{ar} models with hidden semi-Markov states and demonstrate how this algorithm can be used for regime switching, model selection and clustering purposes.
\end{abstract}

%Keywords

\clearpage

%%%%%%%%%%%%%%%%%%%%%%%%%%%
%Chapter - Comments

\section{Introduction} \label{sec:hsmm_Introduction}

Discrete State Space Models (SSM) provide a flexible class of models with applications in ecology, economics, finance, robotics and signal processing \citep{Bulla06, Lindsten13, Papaspiliopoulos20, corenflos21}, among others. They can handle structural breaks, shifts, or time-varying parameters and still have an interpretable structure. Moreover, such models are generative and allow for multi-step forecasting. 
However, analytical forms of the likelihood function are only available in special cases, and standard parameter optimization routines are often challenging to implement. 
%The major challenge in the estimation of \acrshort{ssm}s is thus the generally intractable likelihood function for observations $e_{1:T}=(e_1,\dots,e_T)$, denoted by $p_{\theta}(e_{1:T})$, with $\theta$ denoting parameters, which integrates over the latent state trajectory $s_{1:T}$, such that $p_{\theta}(e_{1:T}) = \int p_{\theta}(e_{1:T}, s_{1:T}) d s_{1:T}$.
Given the observed data $e_{1:T}=(e_1,\dots,e_T)$ and parameter $\theta$, the major challenge in the estimation of \acrshort{ssm}s is thus the generally intractable likelihood function $p_{\theta}(e_{1:T})$, which integrates over the latent state trajectory $s_{1:T}$ such that $p_{\theta}(e_{1:T}) = \int p_{\theta}(e_{1:T}, s_{1:T}) d s_{1:T}$.

%w Throughout this paper, $s_t$ is a latent (hidden) variable at time $t$ while $e_t$ denotes the observed data. Model parameters are denoted with $\theta$. 

A flexible discrete \acrshort{ssm} on which we focus in this paper is known as \acrlong{hsmm}. \acrshort{hsmm}s have a flexible state duration distribution, well suited for processes that remain in any particular state for an extended period of time, and can be considered as generalizations of the well-known basic \acrlong{hmm} introduced in \citet{Baum66}. 
\acrshort{hsmm}s have been employed in ecology, epidemiology, finance \citep{Bulla06, pohle21, visani21} and many other fields \citep{Yu16}, but are typically used more sporadically than their standard \acrshort{hmm} counterparts because the likelihood function is significantly more costly to evaluate. 
In the \acrshort{hmm} case, the likelihood has computational complexity of $\mathcal{O}(K^2T)$, where K = number of latent states, T = number of data points, see \cite{Baum66}. For the \acrshort{hsmm}, this is a much more expensive operation of order $\mathcal{O}(K^2(d_{max} - d_{min})^2T)$ \citep{Murphy02, Wood12}, where $d_{min}$ and $d_{max}$ denote the minimal and maximal state duration in a latent regime. In practice, $(d_{max} - d_{min}) >> K$, as described in more detail in Section \ref{sec:hsmm_HSMM}, which often leads to computationally expensive inference algorithms. Such considerations have led to the use of approximate methods in applications where \acrshort{hsmm}s provide valuable models, see for example \cite{HadjAmar22} and \cite{Xiao18}. 

In this paper, we follow an alternative route aiming to construct efficient computational schemes that operate on the joint space of latent states and parameters using \acrfull{smc} methods that are exact, in the sense that they are only amenable to Monte Carlo error. The fundamental building block of the proposed schemes is the \acrfull{pf}, see for example \cite{Doucet11} and the references therein. 
%, and a particle filter provides significant other advantages in a sequential setting, see chapters \ref{sec:hsmm_HSMM} and \ref{sec:hsmm_BayesianInference}.
Traditionally, \acrshort{smc} samplers such as \acrshort{pf}s have been used to estimate the underlying state sequence of \acrshort{ssm}s, while standard \acrfull{mcmc} samplers facilitate Bayesian inference for the model parameters. 
More recently, combining these methods is becoming increasingly popular, see \citet{daviet18} and \citet{buchholz2020}. 
A natural computational framework that jointly infers the latent state sequence and model parameter is known as \acrfull{pmcmc} \citep{Andrieu10, andrieu2009}. To our knowledge, \acrshort{pmcmc} has not been used for \acrshort{hsmm}s, so we work within this framework aiming to construct an efficient implementation. 
In particular, we focus on the \acrfull{pgibbs} version to implement parameter updates, conditional on the latent state trajectory, via Hamiltonian \acrshort{mcmc} \citep{neal12} variants. 
%We also expand on \acrshort{pmcmc}, noting that it is a batch estimation method, i.e. it will have to be rerun from scratch each time new data become available. 
Given that \acrshort{ssm}s are typically used in applications with data of sequential nature, it is essential to explore techniques where previous parameter estimates can be reused once the data is updated. An example for that is provided by the \acrfull{smc2} algorithm, introduced in \citet{chopin2012}, which can be viewed as an extension of the main \acrshort{smc} framework of \citet{Chopin02} and \cite{DelMoral06}; see also \cite{Dai20} for some recent work that includes a survey of applications in different contexts. 
%The algorithm of \acrshort{smc2} may also be viewed as an application of the pseudo-marginal framework of \cite{andrieu2009} that utilizes the \acrfull{pmcmc} scheme. 
Other similar approaches include \citet{fearnhead10} and \citet{crisan17}. More information on these methods is provided in the Section \ref{sec:hsmm_BayesianInference}.

The major motivation of this paper is thus to develop methods to estimate \acrshort{hsmm}s (1) in a computationally feasible time and (2) in a sequential manner. The contribution of this paper is two-fold: First, we offer \acrlong{smc} schemes on \acrlong{hsmm}s by tailoring ideas from \cite{Andrieu10} and \cite{chopin2013smc2} for batch and sequential estimation. This offers several benefits over standard deterministic filtering techniques, including computational efficiency. 
%To our knowledge this is the first application of such methods in this context and can offer several benefits over standard deterministic filtering techniques including computational efficiency. 
%We explore the arsenal of such methods to tailor them for \acrshort{hsmm}s and develop a Particle Gibbs scheme, where Hamiltonian MCMC \cite{neal12} is used for the parameter updates, and embed this in the environment of \cite{chopin2013smc2}. 
The developed \acrshort{smc} schemes can also facilitate Bayesian model choice and assessment of predictive performance in an efficient manner. 
Second, we propose a novel class of models by linking \acrlong{ar}-type models with \acrshort{hsmm}s to better describe data consisting of financial and econometric time series. 
%Such models may also be viewed as regime-switching with each hidden state corresponding to a regime. 
Sequential estimation of such models is particularly important as \acrshort{ar} \acrshort{hsmm}s have the potential to detect substantial changes in the data, which we illustrate in a case study on data that evolves rapidly during the Covid-19 pandemic.

The paper is organized as follows: Section \ref{sec:hsmm_HSMM} formally introduces \acrshort{hsmm}s via a suitable formulation to apply sequential Monte Carlo methods such as \acrlong{pf}ing. It also provides justification for the use of \acrshort{pf}s instead of deterministic filtering techniques.
In Section \ref{sec:hsmm_BayesianInference}, the developed methodology of this paper is presented, which includes the model choice criteria available from by-products of the estimation process. Section \ref{sec:hsmm_Experiments} explores the performance of the developed methods via simulation based experiments. 
In Section \ref{sec:hsmm_Applications}, we focus on the performance of \acrshort{hsmm}s, estimated with the developed methodology of this paper, on real-world applications such as financial time series of the VIX index. Comparisons of different \acrshort{hsmm}s as well as benchmark \acrshort{hmm}s are conducted. Model selection, and in particular choice of the number of states using \acrshort{smc2}, is also put into test. 
Finally, Section \ref{sec:hsmm_Conclusion} concludes with some relevant discussion.

%All models and algorithm routines in this paper are extensively validated on simulated and real world data. While there are an increasing amount of excellent \acrshort{mcmc} software solutions, such as \citet{stan21, ge18, Salvatier16}, available, we fill a niche by focusing on sequential estimation and providing full support for \acrshort{ssm}s with arbitrary model dynamics.

%The remainder of this paper is structured as follows: chapter 2 introduces the Bayesian inference techniques used in this paper. Chapter 3 discusses HSMMs and their advantages. Chapter 4 specifies our SMC2 tuning and kernel settings, and discusses several model selection criteria that come as by-product from the estimation process. \textcolor{red}{In Chapter 5 , Experiments ( do we even need that?)...} Chapter 6 is the application section, where we show the usability of HSMMs on financial data. The paper ends on chapter 7 with our concluding thoughts.

%%%%%%%%%%%%%%%%%%%%%%%%%%%
%Chapter - Comments

\section{Hidden Semi-Markov Model} \label{sec:hsmm_HSMM}

A standard \acrlong{hmm} may be specified via a bivariate stochastic process $ \{ e_t, s_t \}_{t = 1,2,\ldots} $, where $s_t$ is an unobserved Markov chain and $e_t$ is an observed sequence of independent random variables, conditional on $s_t$. The model is fully specified by the transition distribution $f_\theta$, $s_t \sim f_\theta(s_t \mid s_{t-1})$, ~$t \geq 2$, the corresponding initial distribution $\pi_\theta$, $s_1 \sim \pi_\theta( s_1 )$, and the observation distribution $g_\theta$, $e_t \sim g_\theta(e_t \mid s_{t})$, ~$t \geq 1$. Directly computing the likelihood function of this model involves summing up over all possible state sequences, 
\begin{equation} 
p(e_{1:T}) = \sum_{s_{1:T}} p( e_{1} \mid s_{1}) p( s_{1}) \prod_{t=2}^T p( e_{t} \mid s_{t}) p( s_{t} \mid s_{t-1}).
\label{eq:hsmm_Chp3_HMM_likelihood}
\end{equation}
Hence, various filtering techniques have been proposed that take into account the memory of the latent state variable to reduce the computational costs to $\mathcal{O}(K^2T)$. 
One shortcoming of \acrshort{hmm}s is their explicit distributional assumption regarding the duration in any particular state. To give insight into this issue, we denote $p( s_{t+k} = j,~ s_{t+1:t+k-1} = i \mid s_{t} = i )$, the probability a state remains in any current state until it switches, as state duration distribution. In the \acrshort{hmm} case, this probability is implicitly geometric. Set $p(s_t = i \mid s_{t-1} = i ) = \mathcal{T}_{ii}$, and assume there are only 2 states, then for a homogeneous Markov chain, using the chain rule and the Markov assumption, it holds: 
\begin{equation}
\begin{split} 
p( s_{t+3} = j, s_{t+2} = i, s_{t+1} = i \mid s_{t} = i ) &= p( s_{t+3} = j\mid s_{t+2} = i) p(s_{t+2} = i,  \mid s_{t+1} = i) p( s_{t+1} = i \mid s_{t} = i ) \\
&= (1 - \mathcal{T}_{ii}) * \mathcal{T}_{ii}^2
\end{split}
\label{eq:HMM_geom1}
\end{equation}

\begin{figure}[htp]
\centering
\begin{subfigure}[b]{0.5\textwidth}
    \includegraphics[width=1\textwidth]{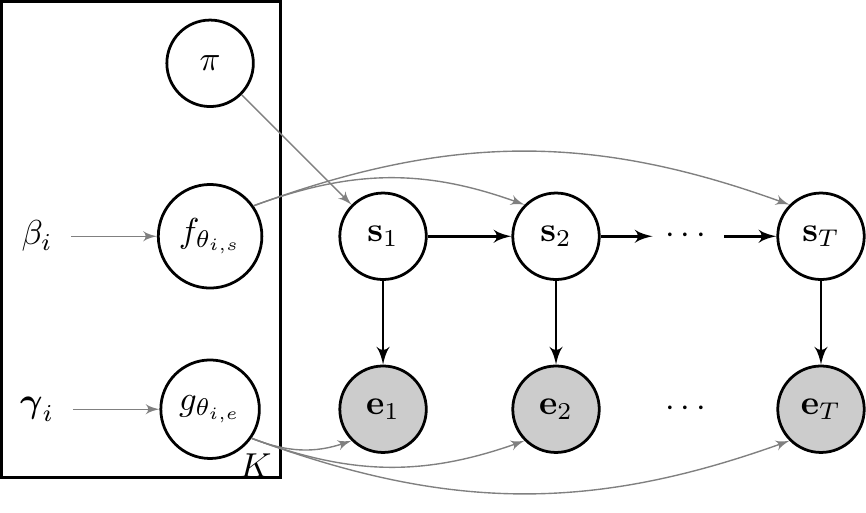}
    %\caption{$K$-state Bayesian HMM, parameter $\theta$ and hyper-parameter $\{ \beta, \gamma\}$. The shaded nodes $e_t$ denote the observed data at time $t$, while the unshaded nodes indicate the latent state $s_t$. $\theta_{i, s}$ denotes the parameter at state $i$ for latent state $s$, and $\beta_i$ the corresponding hyper-parameter. $f$ is the transition distribution $s$, $g$ is the observation distribution for $e$. $\pi$ denotes the initial distribution for $s$.}
    \caption{}
	\label{fig:hsmm_Chp2_BayesianHMM}
\end{subfigure}
\hfill %uncomment if want to have figures side by side (only if textwidth small enough
\begin{subfigure}[b]{0.5\textwidth}
   \includegraphics[width=1\textwidth]{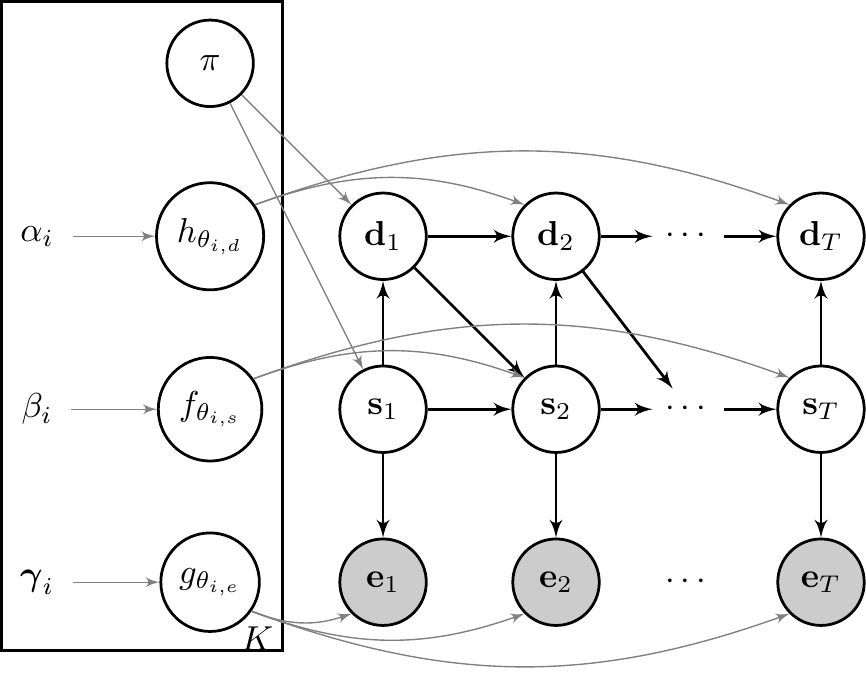}
   % \caption{$K$-state Bayesian HSMM, parameter $\theta$ and hyper-parameter $\{ \alpha, \beta, \gamma \}$. The shaded nodes $e_t$ denote the observed data at time $t$, while the unshaded nodes indicate latent duration $d_t$ and state $s_t$. $\theta_{i, d}$ denotes the parameter at state $i$ for latent duration $d$, and $\alpha_i$ the corresponding hyper-parameter. $h$ and $f$ are the transition distributions for $d$ and $s$, $g$ is the observation distribution for $e$. $\pi$ denotes the initial distribution for $d$ and $s$.}
    \caption{}
    \label{fig:hsmm_Chp2_BayesianHSMM}
\end{subfigure}
\caption{Figure \ref{fig:hsmm_Chp2_BayesianHMM} depicts a $K$-state Bayesian HMM, parameter $\theta$ and hyper-parameter $\{ \beta, \gamma\}$. The shaded nodes $e_t$ denote the observed data at time $t$, while the unshaded nodes indicate the latent state $s_t$. $\theta_{i, s}$ denotes the parameter at state $i$ for latent state $s$, and $\beta_i$ the corresponding hyper-parameter. $f$ is the transition distribution $s$, $g$ is the observation distribution for $e$. $\pi$ denotes the initial distribution for $s$. Figure \ref{fig:hsmm_Chp2_BayesianHSMM} depicts a$K$-state Bayesian HSMM, parameter $\theta$ and hyper-parameter $\{ \alpha, \beta, \gamma \}$. The shaded nodes $e_t$ denote the observed data at time $t$, while the unshaded nodes indicate latent duration $d_t$ and state $s_t$. $\theta_{i, d}$ denotes the parameter at state $i$ for latent duration $d$, and $\alpha_i$ the corresponding hyper-parameter. $h$ and $f$ are the transition distributions for $d$ and $s$, $g$ is the observation distribution for $e$. $\pi$ denotes the initial distribution for $d$ and $s$.}
\end{figure}

In general, for $t+k$ steps, it holds that
\begin{equation}
\begin{split} 
p( s_{t+k} = j, \dots, s_{t+1} = i \mid s_{t} = i ) &= (1 - \mathcal{T}_{ii}) * \mathcal{T}_{ii}^{k-1} \\
&= Geometric_{ \mathcal{T}_{ii} },
\end{split}
\label{eq:HMM_geom2}
\end{equation}
where the geometric distribution has to be interpreted as the length of state duration up to and including the transition to the other state. For processes that tend to stay in any particular state for a long-time horizon, this may be a poor modelling choice. Alternatively, the state duration could be explicitly modelled. A \acrlong{hsmm}, see \citep{Murphy02, Yu10, Yu16}, is a generalization of an \acrshort{hmm}, which may be viewed as a \acrshort{hsmm} with Geometric state duration distribution. 
A graph structure and a comparison to the standard \acrshort{hmm} can be seen in figures \ref{fig:hsmm_Chp2_BayesianHMM} and \ref{fig:hsmm_Chp2_BayesianHSMM}. 
A specific formulation of the \acrshort{hsmm} that explicitly defines the duration distribution is known as \acrfull{edhmm}. Transitions are allowed only at the end of each state, resulting in the following definition:
\begin{definition}{Hidden semi-Markov Model (HSMM) } \label{def:HSMM}
	A hidden semi-Markov model is a bivariate stochastic process $ \{ e_t, z_t \}_{t = 1,2,\ldots} $, where $z_t = \{ s_t, d_t \}$ is an unobserved semi-Markov chain and, conditional on $z_t$, $e_t$ is an observed sequence of independent random variables. The model is fully specified by the transition distribution $f_\theta( s_t \mid s_{t-1}, d_{t-1} )$ of $s_t$
	\begin{equation} 
    s_t \sim  \begin{cases} %\mid \{ S_{t-1} = s_{t-1}, D_{t-1} = d_{t-1} \} 
    \delta( s_{t}, s_{t-1}) &\text{ $d_{t-1} > 0$ }\\
    f_\theta( s_t \mid s_{t-1}, d_{t-1} ) &\text{ $d_{t-1} = 0$ },
    \end{cases}
    \label{eq:hsmm_Chp3_EDHMM_transition}
    \end{equation}
    the duration distribution $h_{\theta}$ of $d_t$
    \begin{equation} 
    d_t \sim \begin{cases} %\mid \{ S_{t} = s_{t}, D_{t-1} = d_{t-1} \} 
    \delta( d_{t}, d_{t-1} - 1) &\text{ $d_{t-1} > 0$ }\\
    h_\theta( d_t \mid s_{t}, d_{t-1}) &\text{ $d_{t-1} = 0$ },
    \end{cases}
    \label{eq:hsmm_Chp3_EDHMM_duration}
    \end{equation}
    
    the corresponding initial distribution $\pi_{\theta}$ of $z_t$, and the observation distribution $g_{\theta}$, $e_t \sim g_{\theta}(e_t \mid s_t)$,
    \begin{equation} 
    e_t \sim g_\theta(e_t \mid s_{t}). %\mid \{ S_{t} = s_{t} \} 
    \label{eq:hsmm_Chp3_EDHMM_observation}
    \end{equation}
    where $\delta(a,b)$ is and indicator function and equals $1$ if $ a = b$ and $0$ otherwise.
\end{definition}

Popular choices for the duration distribution $h_\theta$
%, other than the Geometric distribution which corresponds to the \acrshort{hmm} case, 
are the Poisson or the Negative Binomial distribution, for greater flexibility at the cost of an additional model parameter per state. 
The observation distribution $g_{\theta}$ can be set according to the specifics of the application at hand, which includes higher order data dependency such as $g_{\theta}(e_t \mid s_t, e_{t-k:t-1})$, for $t \ge 1$. 
An example for an excellent data dependency use case is provided in Section \ref{sec:ApplicationsVIX}, where \acrshort{ar}(1) models are used in each latent regime.
%with specifications based on the AR(1) that allow for hidden states, potentially corresponding to different market regimes.
The joint distribution of an \acrshort{edhmm} given the parameter can be stated as
\begin{equation} 
p_\theta(s_{1:T}, d_{1:T}, e_{1:T}) = \pi_\theta(s_1) \pi_\theta(d_1) g_\theta( e_1 \mid s_{1}) \prod_{t = 2}^{T} f_\theta(s_t \mid s_{t-1}, d_{t-1}) h_\theta( d_t \mid s_{t}, d_{t-1}) g_\theta( e_t \mid s_t)  %S_{1:T} = s_{1:T}, D_{1:T} = d_{1:T}, E_{1:T} = e_{1:T}
\label{eq:hsmm_Chp3_EDHMM_Joint}
\end{equation}

The likelihood can be obtained by integrating out both $s_{1:T}$ and $d_{1:T}$. Due to the additional latent variables $d_{1:T}$, this is a much more computationally expensive operation than in the standard HMM case. In order to gain more insight on this, we can shrink the graphical model structure of a \acrshort{hmm} and \acrshort{hsmm} to a single time step. In order to compute the likelihood of observation $e_{t+1}$ given the current state $s_t$, a sum over all possible state transitions has to be taken, as can be seen in equation \eqref{eq:HMM_SeqMix}. This resembles a standard mixture model computation, with the transition matrix of the \acrshort{hmm} replacing the mixture component weights.
\begin{equation}
\begin{split} 
	p( e_{t+1} \mid s_{t} = k) &= \sum_{s_{t+1}} p( e_{t+1}, s_{t+1} \mid s_{t} = k) \\
	&= \sum_{s_{t+1}} p( s_{t+1} \mid s_{t} = k) p( e_{t+1} \mid s_{t+1}) \\
\end{split}
\label{eq:HMM_SeqMix}
\end{equation}

For the \acrshort{hsmm}, however, the duration variable means an additional sum over a random variable that has at worst an infinite number of terms in the case of duration distributions with countably infinite support, shown in equation \eqref{eq:HSMM_SeqMix}.

\begin{equation}
\begin{split} 
	p( e_{t+1} \mid s_{t} = k, d_{t} = j) &= \sum_{s_{t+1}} \sum_{d_{t+1}} p( e_{t+1}, s_{t+1}, d_{t+1} \mid s_{t} = k, d_{t} = j) \\
	&= \sum_{s_{t+1}} \sum_{d_{t+1}} p( s_{t+1} \mid s_{t} = k, d_{t} = j) p( d_{t+1} \mid s_{t+1}, d_{t} = j) p( e_{t+1} \mid s_{t+1}) \\
\end{split}
\label{eq:HSMM_SeqMix}
\end{equation}

Note that $\sum_{z_{t+1}} = \sum_{d_{t+1}}\sum_{s_{t+1}}$, which sums up all possible durations over all states, has at worst an infinite number of terms if the duration distributions have countably infinite support, and at best a large number of terms for long sequences, see~\citet{Wood12}. The standard approach to tackle this problem is to set up a minimum and maximum duration $d_{min}$ and $d_{max}$, where the computational complexity of the forward-backward algorithm reduces to $O(T(K(d_{max} - d_{min})^2) $, compared to the original $O(TK^2)$ in the \acrshort{hmm}, see \citet{Murphy02}. Choosing an appropriate maximum duration varies depending on the underlying data. If the truncation is too small, then inference will typically fail, if is too large then calculations might become infeasible. Hence, $(d_{max} - d_{min})$ may increase the computational complexity to burdensome levels, which requires the modeler to set $d_{max}$ too small. Other approaches include \citep{Johnson13, Johnson14}, who decrease computational complexity by censoring the initial or end time. To give a numerical example, in order to appropriately estimate a \acrshort{hsmm}, we assume that at least a single state transition has to occur, hence $d_{max} < T$, but $(d_{max} - d_{min}) >> K$. The computational costs for a $5$-state \acrshort{hmm} and $1000$ data points would be $\mathcal{O}(5^2 \times 1000)$, while for the \acrshort{hsmm}, assuming $d_{max} = 500$ and $d_{min} = 0$, $\mathcal{O}(5^2 \times 500^2 \times 1000)$ for a single likelihood call. 

Alternatively, a particle filter can be used for the likelihood computation in $\mathcal{O}(NT)$ operations, even if the model has \acrshort{hsmm} dynamics.
Exact inference is retained, subject to Monte Carlo error, using the \acrlong{pmcmc} algorithm \citep{Andrieu10}. 
$N$ denotes the number of particles used in the filter, and we observed that it is sufficient to set $N = \frac{T}{2}$ for sequential Monte Carlo schemes on \acrshort{hsmm}s, see Section \ref{sec:hsmm_BayesianInference} for more detail. 
 %It is a priori unclear if $\mathcal{O}(NT)$ is drastically smaller than $\mathcal{O}(K^2(d_{max} - d_{min})^2T)$ for small K. 
%To get a feel of potential computational improvements, as can be seen in Section \ref{sec:hsmm_BayesianInference} a choice of  
%Consequently, particle filter methods may be better suited than deterministic filtering techniques for \acrshort{hsmm}s. Finally, another potential major advantage of particle filter methods is that they can provide fully online algorithms in cases where the parameters are fixed based on previous historical data analyses.

%%%%%%%%%%%%%%%%%%%%%%%%%%%
%Chapter - Comments

\section{Bayesian Inference on Hidden Semi-Markov Models} \label{sec:hsmm_BayesianInference}

In a Bayesian framework, the typical goal is to infer the posterior distribution of the model parameter $\theta$ given the observed data $e_{1:T}$, $p(\theta \mid e_{1:T}) = \frac{ p_\theta(e_{1:T}) ~ p(\theta)}{ p( e_{1:T} )}$. This is a challenging task for \acrshort{ssm}s as it involves integrating $s_{1:T}$ over the likelihood $p_\theta(e_{1:t} ) = \int p_\theta(e_{1:T}, s_{1:T})~d s_{1:T}$, which is typically intractable or costly to evaluate. Hence, usually the full posterior distribution
\begin{equation}
    p(s_{1:T}, \theta \mid e_{1:T}) = \frac{ p_\theta(e_{1:T} \mid s_{1:T}) ~ p_\theta(s_{1:T}) ~ p(\theta)}{ p( e_{1:T} )}
    \label{eq:hsmm_Chp2_FullPosterior}
\end{equation} 
is inferred.
If $s_{1:T}$ is continuous, the target distribution  
$p( s_{1:t}, \theta \mid e_{1:t}) \propto p(e_{1:t} \mid s_{1:t}, \theta ) \times p( s_{1:t}\mid \theta ) \times p( \theta ) $ can theoretically be estimated via \acrshort{mcmc}. However, in this case, a state trajectory of ${p( s_{1:T} \mid {\theta})}$ has to be sampled while evaluating the target function, usually resulting in very poor outcomes of this strategy. 
Alternatively, classic Gibbs sampling strategies could be applied, which iterate the estimation process between sampling the latent states given the continuous model parameter and vice versa. 
However, for the latent state trajectory proposal step, a forward-backward algorithm would have to be employed again, as other choices such as one-at-a-time updates or overlapping blocks are known to cause slow mixing \citep{Kalogeropoulos10, Golightly09} of the Markov chain. 

A more general attempt to jointly target the full joint posterior $p( s_{1:T}, \theta \mid e_{1:T}) $ can be shown as follows:
\begin{itemize}
	\item 1. propose $\theta^{\star} \sim f(\theta^{\star} \mid \theta)$ and  $s^{\star}_{1:T} \sim p_{\theta^{\star}}(s^{\star}_{1:T} \mid e_{1:T})$,
	\item 2. accept $(\theta^{\star}, s^{\star}_{1:T})$ with acceptance probability
	\begin{equation} 
	a( (s^{\star}_{1:T}, \theta^{\star}), (s_{1:T}, \theta) ) = \frac{p_{\theta^{\star}}(e_{1:T})}{p_\theta(e_{1:T})} \frac{p(\theta^{\star})}{P(\theta)} \frac{q(\theta \mid \theta^{\star})}{q(\theta^{\star} \mid \theta)}.
	\label{eq:hsmm_Chp2_PMCMC}
	\end{equation}
\end{itemize}

The last term in equation \eqref{eq:hsmm_Chp2_PMCMC} has been simplified by using the \acrfull{bmi} of \citet{Chib95}. This framework allows to jointly sample $\theta$ and $s_{1:t}$, but the in general intractable likelihood function is still contained in step 3.
While this term can be computed analytically for the discrete \acrshort{hsmm}
via the so-called forward-backward algorithms, they are prohibitively expensive to run, 
as described in Section \ref{sec:hsmm_HSMM}.
Going forward, we introduce the algorithmic machinery known as \acrlong{pmcmc} \citep{Andrieu10},  which replaces the likelihood evaluation with an estimate $\hat{p}_\theta(e_{1:T})$ from a \acrlong{pf}. %, thereby circumventing the need to analytically evaluate the likelihood function of the \acrlong{ssm} of choice. 

\begin{comment}
Alternative schemes for likelihood evaluation use beam sampling, see \citet{Wood12}, or define custom Gibbs sampler, see \citet{Johnson13} in this step. However, the particle filter usage has major advantages in a sequential estimation context and can be updated online during major parts in the \acrshort{smc2} framework, see section \ref{subsec:hsmm_PF} and \ref{subsec:hsmm_SMC2}.
\end{comment}

%%%%%%%%%%%%%%%%%%%%%%%%%%%%%%%%%%%%%%%%%%%%%%%%%%%%%%%%%%%%%%%%%%%%%%%%%%%%%%%%%%%%%%%%%%%%%%%%%%%%%%%%%
\subsection{Particle Filtering} \label{subsec:hsmm_PF}
\acrlong{pf}s are often used to solve filtering equations in the form of $\pi_t( x_{1:t} ) = \frac{ \tau_t(x_{1:t}) }{ z_t }$. The goal is to sequentially sample a sequence of random variables, $x_t, t \in (1,..., T)$ that come from a sequence of target probabilities $\pi_t( x_{1:t} )$ with the same computational complexity at each time step. If it impossible to directly sample from $\pi_t$, a similar proposal distribution $q_t$ can be used, s.t. $\pi_t(x_{1:t}) > 0 \Rightarrow q_t(x_{1:t}) > 0 $. 
The fraction of $\tau_t(x_{1:t})$ and $q_t(x_{1:t})$ is known as un-normalized weight function $ w_t( x_{1:t} ) = \frac{ \tau_t(x_{1:t}) }{ q_t(x_{1:t}) }$, s.t. the target distribution can be rewritten as $ \pi_t(x_{1:t}) = \frac{w_t(x_{1:t}) q_t(x_{1:t})}{z_t}$. 
This method is recognized as \acrfull{is}. Often, the variable of interest is the normalizing constant $z_t= \int \tau_t(x_{1:t}) d x_{1:t} = \int w_t(x_{1:t}) q_t(x_{1:t}) d x_{1:t} $, which can be approximated via the un-normalized weight functions
\begin{equation}
\hat{z_t} = \frac{1}{t} \sum_{i=1}^{K} \frac{\tau_t(x_{1:t})}{q_t(x_{1:t})} = \frac{1}{t} \sum_{i=1}^{K} w_t(x^i_{1:t}). 
\end{equation}

This technique does, unfortunately, rapidly degenerate as $t$ becomes larger. A technique to sequentially sample from such distributions is called \acrfull{sis}, which keeps the computational costs fixed given additional time steps by decomposing the joint distribution as $\tau_t(x_{1:t}) = \tau_{t-1}(x_{1:t-1}) \tau_t( x_t \mid x_{1:t-1})$. Similarly, the importance distribution can be decomposed as
%\begin{equation}
$q_t( x_{1:t} ) = q_1(x_1) \prod_{n=2}^{t} q_n(x_n \mid x_{1:n-1})$. 
%\end{equation}
%only up to $t-1$, so for each additional time step t you can just sample $q_t(x_t \mid x_{t-1})$ at fixed computational costs, and then apply the equations above to get the desired quantity. 
The associated un-normalized weights can then be computed recursively via
%\begin{equation}
$w_t( x_{1:t} ) =  w_1(x_1) \prod_{k=1}^{t} \alpha_k(x_{1:k})$,
%\end{equation}
where the incremental importance weights $\alpha_t(x_{1:t})$ are given as $ \alpha_t(x_{1:t}) = \frac{ \tau_t(x_{t} \mid x_{1:t-1}) }{ q_t(x_t \mid x_{1:t-1}) }$. 

In most state space models, the memory for $q_t(x_t \mid x_{1:t-1})$ is limited, so the target distribution can be evaluated and sampled from via the methods above at fixed computational costs.
The only freedom in this framework is choosing an appropriate $q_t$. 
%Note that in sequential problems, this formulation would let the computational time complexity grow with $t$, but all the most common \acrshort{ssm}s have limited memory, hence the computational costs remain fixed per additional time step. 
%For a given $\theta$, \acrshort{sis} algorithms approximate $\pi_{\theta}(x_{1:t})$ by propagating forward a set of particles over time according to the proposal $q$, with a fixed computational costs according to the memory properties of the corresponding state space model. 
Unfortunately, it can be shown that the variance of the corresponding weights grows with $t$, and we refer to this problem as weight degeneracy. Due to weights degeneracy, the variance for the estimator of the normalizing constant $\hat{z}$ is also increasing with $t$. 
To alleviate this obstacle, a resampling step for the particle trajectories $x_{1:t}$ that normalizes the corresponding weights can be applied. Algorithms that do so at each iteration are known as \acrfull{sir}. 
However, this creates a new challenge known as sample path degeneracy, which refers to the problem that continuously resampling particle paths ultimately ends with very few unique trajectories. 
Balancing weight and sample path degeneracy is an ongoing research topic, and the most common method is to resample trajectories only at specific iterations, for example, if the \acrfull{ess}
\begin{equation}
    ESS_t = \frac{1}{\sum_{n=1}^N \left( \frac{w_{t}( x^n_{1:t} )}{\sum_{i=1}^N w_{t}( x^i_{1:t})} \right)^2}    
\end{equation}
of the particles is less than an a priori set threshold. For a discussion on several different resampling techniques, see \citet{Douc05}. Adaptive resampling mitigates the exploding variance of the particle weights and keeps sample path degeneracy in check.
Algorithms that apply this machinery are commonly referred as \acrlong{pf}s. 
%They propagate forward a set of particles through the underlying data by weighting and adaptively resampling particle trajectories. 
They have a fixed computational complexity that is both linear in time $T$ and in number of particles $N$, $\mathcal{O}(NT)$, 
and return an estimate of the normalizing constant $\hat{z}_t$ and a particle path of $\hat{\pi}(x_{1:t})$.
%While $\hat{Z}_t$ is only an estimate, the computational costs are fixed for all state space models used in this paper, even if the term may not be evaluated analytically. 
In the \acrshort{ssm} case, the joint distribution and the normalizing constant are of the form $\tau_t(x_{1:t}) = p_\theta(s_{1:t}, e_{1:t})$ and $z_t = p_\theta(e_{1:t})$. An approximation for the likelihood can be computed via the weights $\hat{z_t} = \frac{1}{N} \sum_{n=1}^{N} w_t( s_{1:t}^n, e_{1:t} )$ , which can be decomposed in the following recursive form:
\begin{equation}
w_t( s_{1:t}, e_{1:t} ) = w_{1}( s_{1}, e_{1} ) ~ \prod_{k=1}^{t} \alpha_k(s_{1:k}, e_{1:k}).
\end{equation}

The incremental weight $\alpha$ is defined as $\alpha_t(s_{1:t}, e_{1:t}) = \frac{ p_\theta( e_{t} \mid s_{1:t}, e_{1:t-1} ) ~p_\theta( s_{t} \mid s_{1:t-1}, e_{1:t-1} ) }{ q(s_t \mid s_{1:t-1}, e_{1:t}) }$, and the full likelihood estimate can be expressed as $\hat{z_T} = \frac{1}{N} \sum_{n=1}^{N} \prod_{k=1}^{T} \alpha_k(s^n_{1:k}, e_{1:k})$, which is usually the preferred method in particle filter software implementations as this avoids memory allocations. This permits the estimation of the incremental likelihood $p_{\theta}(e_t \mid e_{1:t-1}) \approx \frac{1}{N} \sum_{n=1}^{N} \alpha_t(s^n_{1:t}, e_{1:t})$ as well, which becomes relevant in the model selection Section \ref{subsec:hsmm_ModelSelection}. 
$p_\theta( e_{t} \mid s_{1:t}, e_{1:t-1} )$ and $p_\theta( s_{t} \mid s_{1:t-1}, e_{1:t-1} )$ are model distributions, and have usually limited memory. The only free distribution to choose is $q(s_t \mid s_{1:t-1}, e_{1:t})$, which should ideally look like $p_\theta(s_t \mid s_{1:t-1}, e_{1:t})$. A common and simple choice is known as \acrfull{bpf}, which takes $q(s_t \mid s_{1:t-1}, e_{1:t}) = p_\theta( s_{t} \mid s_{1:t-1}, e_{1:t-1} )$, reducing the incremental weight to $\alpha_t(s_{1:t}, e_{1:t}) = p_\theta( e_{t} \mid s_{1:t}, e_{1:t-1} )$. 
Another popular approach is the so called \acrfull{apf} \citep{Pitt99}, which assumes $\alpha_t(s_{1:t}, e_{1:t})$ to be independent of $s_t$ (in the Bootstrap \acrlong{pf}, this does not hold!). 
See \citep{Kantas15, Doucet11} for a more in-depth review. A pseudo algorithm implementation for a standard \acrshort{pf} can be found in Algorithm \ref{alg:ParticleFilter}, where the auxiliary variable $a^i_{t}$ refers to the ancestor path of a particular particle $s^i$ at time t. 
Hence a particle trajectory can be recursively defined as $s^i_{1:t} = (s^{a^i_{t}}_{1:t-1}, s^i_t)$. Resampling the whole particle trajectory is equivalent to sampling a new ancestor path. Note that it is usually much faster to sample ancestors one at a time and then recursively recover the resampled particle path than to resample the whole particle trajectory at each iteration. A variant of this algorithm is known as \acrfull{cpf}, where a single particle path, $s'_{1:T}$, is chosen a priori as reference trajectory. This implementation tracks a slightly different target distribution, $\hat{p}(s_{1:T} \mid s'_{1:T}, e_{1:T})$, and is used in Section \ref{sec:hsmm_Experiments} and \ref{sec:hsmm_Applications}. A pseudo algorithm for this \acrfull{cpf} with ancestor sampling can be found in Algorithm \ref{alg:ConditionalParticleFilter}, and a more thorough review can be read in \citep{lindsten14, lindsten15}.

Once a particle filter has been run, observed and latent data can be forecasted by first sampling a new state $s_{T+1} \sim p_{\theta}(\cdot \mid s_{1:T}, e_{1:T}) $ and then a new data point given this state $e_{T+1} \sim p_{\theta}(\cdot \mid s_{1:T+1}, e_{1:T})$.
$s_{T+1}$ can be sampled by forward propagating algorithm \ref{alg:ParticleFilter} or \ref{alg:ConditionalParticleFilter} from $T$ to $T+1$, thereby reusing particles from $1$ to $T$. This can be repeated for multiple time steps as well, resulting in a very fast procedure to sample from predictive distributions. 
%However, this mechanism will be noisy if the times series becomes too long, and model parameter may change the further the time series is increasing in time t.

\begin{comment}    
If $\theta$ is fixed, it is straightforward to sample from $p_\theta(s_{T+1:T+i}, e_{T+1:T+i} \mid s_{1:T}, e_{1:T})$, for $i \geq 1$, by
\begin{itemize}
	\item (1) running a particle filter that targets $p(s_{1:T} \mid e_{1:T}, \theta)$.
	\item (2) \textbf{for $n \leftarrow T+1$ to $T+i$ do} 
	\begin{itemize}
		\item (i) draw $s_{n} \sim p_\theta(\cdot \mid s_{1:n-1}, e_{1:n-1})$,
		\item (ii) draw $e_{n} \sim p_\theta( \cdot \mid s_{1:n}, e_{1:n-1})$.
	\end{itemize}
\end{itemize}

In (2), the \acrlong{pf} can be updated online to propagate particles up to $T+i$, resulting in a very fast procedure to sample from predictive distributions. However, this mechanism will be noisy if the times series becomes too long, and model parameter may change the further the time series is increasing in time t.
\end{comment}

%%%%%%%%%%%%%%%%%%%%%%%%%%%%%%%%%%%%%%%%%%%%%%%%%%%%%%%%%%%%%%%%%%%%%%%%%%%%%%%%%%%%%%%%%%%%%%%%%%%%%%%%%
\subsection{Particle Markov Chain Monte Carlo}

The most common Bayesian inference technique for model parameter $\theta$ is known as \acrlong{mcmc}. Basic familiarly with this concept is assumed, and we refer to \cite{Rosenthal14} for a more detailed review about standard \acrshort{mcmc} techniques. A pseudo algorithm for a basic Metropolis step can be found in Algorithm \ref{alg:MH}. The major difficulty in algorithm \ref{alg:MH} is finding a good proposal distribution $f$, which often results in slow mixing. A \acrshort{mcmc} kernel that automatically tunes its proposal distribution at the cost of additional tuning hyper-parameter is known as \acrfull{hmc}, see \citep{neal12} for an introduction and \citep{betancourt18} for a review on this topic. A pseudo algorithm can be seen in Algorithm \ref{alg:HMC}. The additional tuning hyper-parameter can be configured on the fly in the famous extension \acrfull{nuts}, which was proposed by \citet{Gelman14}. Notable improvements for this algorithm are suggested in \citet{betancourt16}. %It is out of the scope of this paper to explain all adjustments in detail, and we instead refer to said literature. 

Note that both the \acrshort{hmc} and \acrshort{nuts} kernel require the target density to be fully differentiable with respect to the model parameter $\theta$. In the \acrshort{ssm} case, targetting the marginal posterior distribution $p(\theta \mid e_{1:T}) \propto p_\theta(e_{1:T}) ~p(\theta)$ is difficult or impossible via \acrshort{mcmc}, as the latent variables in $p_\theta(e_{1:T}) = \int_{s_{1:T} } p_\theta(e_{1:T}, s_{1:T})$ have to be integrated out in the proposal ratio. 
However, $p_\theta(e_{1:T}, s_{1:T})$ is usually computable pointwise, so $p(s_{1:T}, \theta \mid e_{1:T})$ can be targeted. In the \acrfull{pmh} case, formally introduced in \citep{Andrieu10} and shown in pseudo Algorithm \ref{alg:PMCMC}, a particle filter is used to obtain approximations for $p_{\theta}(e_{1:T} )$ and $p_{\theta}(s_{1:T} \mid e_{1:T})$ as substitutes for the analytical solutions to target $p(s_{1:T}, \theta \mid e_{1:T})$ jointly. 
In this setting, \citep{andrieu2009} have shown the puzzling result that one can do so and still target the exact posterior distribution of interest. A major difficulty for this method is finding a good \acrshort{mcmc} kernel $ K_{mcmc}(e_{1:T}, \theta )$, because the $\theta$ proposal will be accepted based on the particle filter likelihood estimate, so tuning might be very noisy. Moreover, gradient based \acrshort{mcmc} sampler do not work in this case as $p_\theta(e_{1:T})$ typically cannot be evaluated pointwise. A common critique on \acrshort{pmh} is thus that this algorithm is ill-suited for a higher dimensional model parameter $\theta$. 
A \acrshort{pmcmc} variant that can mitigate this is known as \acrfull{pgas}, see \citep{lindsten14, lindsten15}. A pseudo algorithm is shown in Algorithm \ref{alg:PGibbs}. To account for sampling from an approximation via a particle filter and to preserve the invariance principle, a slightly adjusted $\hat{p}_{\theta^{\star}}(s^{\star}_{1:T} \mid s_{1:T}, e_{1:T})$ distribution is used to sample from the state trajectory. This method does not jointly estimate the state sequence and model parameter, but the state sequence is fixed when the new model parameter $\theta^{\star} \sim p_\theta(\theta^{\star} \mid s_{1:T}, e_{1:T})$ are sampled. 
In this step, more advanced \acrshort{hmc} style \acrshort{mcmc} kernels can be used as
%This step can be easily replaced with one of the \acrshort{mcmc} kernels stated earlier 
$p_{\theta^{\star}}(e_{1:T}\mid s_{1:T})$, which is easy to evaluate, replaces $p_{\theta^{\star}}(e_{1:T} )$ in the acceptance ratio. Hence, more advanced \acrshort{mcmc} kernels, such as the \acrshort{nuts} sampler, can be used to estimate model parameter $\theta$ in the \acrshort{pgas} setting. 

%A notable alternative to \acrshort{pgibbs} techniques are classic Gibbs sampling strategies, which iterate the estimation process between sampling the latent states given the continuous model parameter and vice versa. However, to sample the latent variables, forward-backward algorithm would have to be employed again, as existing alternatives such as one-at-a-time updates or overlapping blocks are known to cause slow mixing \citep{Kalogeropoulos10, Golightly09} of the Markov chain. 

%%%%%%%%%%%%%%%%%%%%%%%%%%%%%%%%%%%%%%%%%%%%%%%%%%%%%%%%%%%%%%%%%%%%%%%%%%%%%%%%%%%%%%%%%%%%%%%%%%%%%%%%%
\subsection{Sequential Monte Carlo Squared} \label{subsec:hsmm_SMC2}

In a times series setting, forecasting is of major relevance. The standard way for prediction in a Bayesian setting is simple: obtain the posterior predictive distribution by integrating out the model parameter $\theta$ and, in the \acrshort{ssm} case, the state trajectories ${s_{1:T}}$,

\begin{equation}
\begin{split}
p(e_{T+1} \mid e_{1:T}) & = \int p(e_{T+1},  s_{T+1}, s_{1:T}, \theta \mid e_{1:T}) ~ d s_{T+1}, s_{1:T}, \theta \\
%& =  \int p(e_{T+1} \mid s_{T+1}, s_{1:T}, \theta, e_{1:T}) p(s_{T+1}, s_{1:T}, \theta \mid e_{1:T}) ~ d s_{T+1}, s_{1:T}, \theta \\
&= \int p_\theta(e_{T+1} \mid s_{T+1}, s_{1:T}, e_{1:T}) ~ p_\theta(s_{T+1} \mid s_{1:T}, e_{1:T}) ~ p(s_{1:T}, \theta \mid e_{1:T})  ~ d s_{T+1}, s_{1:T}, \theta.
\end{split} 
\label{eq:hsmm_Chp4_PosteriorPredictive}
\end{equation}

Once a sample for $p(s_{1:T}, \theta \mid e_{1:T})$ is obtained, the predictive distributions for $ s_{T+1} \mid s_{1:T}, e_{1:T}, \theta $ and $ e_{T+1} \mid s_{T+1}, s_{1:T}, e_{1:T}, \theta $ are trivial to sample from. A major drawback of the \acrshort{pmcmc} machinery is that, even though this algorithm primarily works for models suited to times series settings, it only works as batch estimation. Once additional data is observed, the algorithm needs to be run again to target $p(s_{1:T+1}, \theta \mid e_{1:T+1})$. A method that uses \acrshort{pmcmc} in a sequential setting is known as \acrlong{smc2}, see \citet{chopin2012}. \acrshort{smc} based algorithm often expand in the density region, known as density tempering or annealing, or the data dimension, known as data tempering or annealing, see \citet{gunawan21}. \acrshort{smc2} is a data tempering algorithm and moves a collection of particles that consist of the model parameters and latent states by incrementally adding data to the estimation process. At the beginning, particles are drawn from the prior. Subsequent particles are explored iteratively by using multiple \acrlong{pf} and particle \acrshort{mcmc} sampler. At each iteration, N \acrlong{pf} are used to obtain the incremental likelihood estimates $\hat{p}_{\theta^n}(e_{1:t} \mid e_{1:t-1}) = \hat{Z}_t = \frac{1}{N} \sum_{n=1}^{N} \alpha_{\theta^n}(s^n_{1:t}, e_{1:t})$ and state trajectories $s^{n}_{1:t} \sim \hat{p}_{\theta^{n}}(s^{n}_{1:t} \mid e_{1:t})$ for $n = 1,\dots, N$. If the estimates $\hat{p}_{\theta^n}(e_{1:t} \mid e_{1:t-1})$ are getting too noisy, the particles are jittered via \acrshort{pmcmc}. Note that either the Particle Gibbs or the Particle Metropolis Hastings variant can be chosen for the \acrshort{pmcmc} kernel. The density at the final iteration is the posterior distribution of interest. A pseudo algorithm can be seen in Algorithm \ref{alg:SMC2}.

A powerful feature of the \acrshort{smc2} algorithm is that at each iteration, an unbiased estimate of the incremental marginal likelihood $\hat{p}(e_t \mid e_{1:t-1})$ is obtained at practically no extra costs, see equation \eqref{eq:App_IncrementalMarginalLikelihood}. From here on, it is straight forward to obtain an estimate for the marginal likelihood, $\hat{p}(e_{1:T} ) = \prod_{t=1}^T \hat{p}(e_t \mid e_{1:t-1})$, for model comparison or \acrfull{bf} calculations. Moreover, this machinery is highly parallelizable, a feature that is typically difficult to include in standard \acrshort{mcmc} techniques, and most \acrshort{smc2} iterations can be performed online as the particle filter can be propagated forward if no resampling step has been taken at the previous iteration. Additionally, during the propagation step, samples for $s_{t+1}$ and $e_{t+1}$ for posterior predictive distribution analysis can be obtained at each time index.
%Ideally, particles are jittered more often in the initial iterations, in which the data set is still comparatively small, and updates only occasionally towards the end of the data series when parameter reach the typical set. Hence, in practice, \acrshort{smc2}, while using several \acrshort{pf} and \acrshort{pmcmc} kernels, might be faster than a standard \acrshort{pmcmc} run. 
%Moreover, posterior predictive distributions for $s_{t+1}$ and $e_{t+1}$ and an estimate for marginal likelihood $p(e_{1:t})$ for each $t=1, \dots, T$ can be obtained as a by-product. 

\begin{comment}    
Batch data \acrshort{smc}, on the other hand, begins sampling from a flattened target density by scaling the likelihood term with a scalar $\lambda$, $0 \leq \lambda \leq 1$, known as temperature. The target distribution is then stated as 
\begin{equation}
    p_{\lambda_i}(\theta, s_{1:T} \mid e_{1:T}) \propto p_{\theta}(e_{1:T} \mid s_{1:T})^{\lambda_i} p_{\theta}(s_{1:T}) p(\theta).
\end{equation}
The temperature parameter can be adaptively tuned, see \citet{beskos13}, starting from a value close to 0 until 1, in which case sampling is stopped. This method works particularly well for multimodal models, as the initial target density is flat and several modes may be uncovered more easily than in standard \acrshort{mcmc} procedures. Data and density tempering may even be combined, but we will not cover this method any further in this paper.
\end{comment}

%%%%%%%%%%%%%%%%%%%%%%%%%%%%%%%%%%%%%%%%%%%%%%%%%%%%%%%%%%%%%%%%%%%%%%%%%%%%%%%%%%%%%%%%%%%%%%%%%%%%%%%%%
\subsection{Tuning Configurations} \label{subsec:hsmm_Tuning}

%%%%%%%%%%%
\subsubsection{Particle Filter tuning}

In our setup, we used a bootstrap particle filter with transition distribution equal to the model dynamics as defined in Section \ref{sec:hsmm_HSMM}. 
The particle resampling method was chosen to be systematic, and the resampling threshold for the ESS calculation was set to $75\%$. 
The only free tuning parameter in this case is the number of particles $N$. The higher $N$, the lower the variance for the log likelihood estimate, but the higher the computational costs per \acrshort{pmcmc} iteration. 
We note that ultimately, the \acrlong{pmcmc} samples are drawn from the correct target distribution, only subject to \acrlong{mc} error, independent of $N$. However, the mixing of the \acrshort{mcmc} chains might be slower if less particles are used. In this case, a larger $N$ might lead to better results at a fixed computational time horizon.
Often, people set $N$ equal to the number of data points received, but in practice, significantly less particles may be used. 
As a sanity check, we performed two experiments to provide insight.
First, we computed a likelihood estimate for a range of parameter values for $ \theta = \{ \mu, \sigma, r, \phi \}$ of a 2-state \acrshort{hsmm} for 1000 data points. $\mu$ and $\sigma$ represent the parameter for a Normal distribution given a latent state, $r$ and $\phi$ are the parameter of a Negative Binomial duration distribution. The transition distribution has no unknown parameter in the 2-state \acrshort{hsmm} case, as the diagonal elements of the transition matrix are separately modeled by the duration. The true parameter can be seen as vertical grey lines on each subplot in Figure \ref{fig:hsmm_Chp3_HSMM_ParticleFilterEstimate}, which shows the log likelihood estimate for a range of each parameter that was chosen based on the 95\% \acrfull{ci} of a \acrshort{pmcmc} run in Section \ref{sec:hsmm_Experiments}, keeping all other parameter fixed. It can be seen that the variance is reasonably similar for $N = 500$, $1000$ or $2000$.
Second, we directly examined the particle filter performance during a \acrshort{pmcmc} run for the model defined above. The \acrfull{pacf} plot of the log likelihood \acrshort{pf} estimates for a varying number of particles can be seen in Figure \ref{fig:hsmm_Chp3_HSMM_PACF}.
As there is little difference in the likelihood estimate variance between \acrshort{pf}s with $500$ and $2000$ particles and the \acrshort{pacf} look reasonably similar for the same particle range, we will use $N = \frac{T}{2}$ for our analysis going forward.

%Note that this is different for each individual \acrshort{ssm}, for instance we found that as little as $10\%$ is sufficient for a standard \acrshort{hmm}, but the additional latent variable in the \acrshort{hsmm} causes a higher variance for the log likelihood estimate of the particle filter for the latter model. 
%Another interesting observation of this illustration is that the variance of the likelihood estimate is not constant across the whole parameter range for some parameter, so a more optimized tuning technique to choose the number of particles might lead to better results for both the \acrshort{pmcmc} and \acrshort{smc2} algorithm. 

%%%%%%%%%%%
\subsubsection{PMCMC tuning}

Once a particle filter is designed, only a suitable MCMC kernel has to be chosen. As discussed in Section \ref{sec:hsmm_BayesianInference}, more advanced gradient based \acrshort{mcmc} sampler do not work in the \acrlong{pmh} case, as $p_\theta(e_{1:T})$ typically cannot be evaluated pointwise. Thus, we will use \acrlong{pgas} and target $\theta^{\star} \sim p_\theta(\theta^{\star} \mid s_{1:T}, e_{1:T})$ in the MCMC step. As gradients for $p_{\theta^{\star}}(e_{1:T}\mid s_{1:T})$ can easily be calculated in this case, we choose the NUTS MCMC variant \citep{Gelman14, betancourt16} as MCMC kernel, as other kernels often take significantly more proposal steps to move toward the typical set. This is especially relevant for \acrshort{pmcmc} on our model, as we will run a comparatively expensive particle filter after each \acrshort{mcmc} proposal.

%%%%%%%%%%%
\subsubsection{SMC2 tuning} \label{subsec:hsmm_TuningSMC2}

As described in pseudo-algorithm \ref{alg:SMC2}, \acrshort{smc2} has a particle filter and a \acrshort{pmcmc} algorithm assigned for each particle. 
These particles may be propagated in parallel, so the number of \acrshort{smc2} are typically chosen to be a multiple of the available computer cores.
%so for maximum speedup the number of particles should match the available cores, or be a multiple of them. 
Initial model parameter drawn from the prior distributions, and the associated \acrshort{pf} and \acrshort{pmcmc} algorithm will be initiated based on the starting data of length $t_0<<T$. 
%$t_0$ should be chosen such that initial computation is cheap and covers a broad range of values for each parameter, but enough that that different states are observed for initial kernel tuning. Once the initial data points have been determined, the \acrshort{pf} and \acrshort{pmcmc} pairs for each particle will have a few test runs before the \acrshort{smc2} propagation begins. This slightly changes the model evidence estimate, see Section \ref{subsec:hsmm_ModelSelection}, but greatly improves initial model parameter estimates and simplifies tuning procedures for the individual \acrshort{pf} and \acrshort{pmcmc} kernels. 
Another tuning parameter is the number of jittering steps in the resampling step. As a guideline, we will continue jittering until the maximum parameter correlation is below $75\%$. 

%%%%%%%%%%%%%%%%%%%%%%%%%%%%%%%%%%%%%%%%%%%%%%%%%%%%%%%%%%%%%%%%%%%%%%%%%%%%%%%%%%%%%%%%%%%%%%%%%%%%%%%%%
\subsection{Model Selection} \label{subsec:hsmm_ModelSelection}

Once parameter are estimated, how should the performance of a model be evaluated? A powerful method for comparison is to validate models based on their marginal likelihood $p(e_{1:T} ) = \int p(e_{1:T}, \theta)~d \theta = \prod_{t=1}^T p(e_t \mid e_{1:t-1})$. This distribution is typically intractable or very costly to evaluate, but during an \acrshort{smc2} run, an estimate for $p(e_t \mid e_{1:t-1})$ can be obtained at practically no extra cost at each time step. In the particle filter propagation step in algorithm \ref{alg:SMC2}, the incremental weight given current model parameter $\theta_m$ is computed via  
\begin{equation}
    \hat{p}_{\theta_n}(e_t \mid e_{1:t-1}) = \frac{1}{M} \sum_{m=1}^M \alpha_{t, \theta_n}(s^m_{1:t}, e_{1:t}),
    \label{eq:hsmm_increments}
\end{equation}
where $\alpha_{t, \theta_n}$ is defined as in Section \ref{subsec:hsmm_PF} and $M$ is the number of particles that are used for the particle filter associated to continuous parameter vector $\theta_n$. After all particles have been propagated forward, a Monte Carlo estimator for $p(e_t \mid e_{1:t-1})$ is obtained by weighting these likelihood increments from equation \eqref{eq:hsmm_increments} with the corresponding normalized particle weight $w_n \propto w_{n-1} \hat{p}_{\theta_n}(e_t \mid e_{1:t-1})$ associated to parameter $\theta_n$,
\begin{equation}
    \hat{p}(e_t \mid e_{1:t-1}) = \sum_n w_n \times \hat{p}_{\theta_n}(e_t \mid e_{1:t-1}).
\end{equation}
Moving forward, we refer to $p(e_{t+1} \mid e_{1:t})$ as (one step ahead) \acrfull{pl} at $t+1$, $PL_{t+1}$, see \cite{Kastner16}. After the final iteration, the marginal likelihood estimate can then be computed as 
$\hat{p}(e_{1:t}) = \prod_{i=1}^t \hat{PL}_t$.
Based on the cumulative sums of log \acrshort{pl}s, 
model choice can be performed via the so called \acrfull{clpbf}. 
To compare model $A$ and $B$, for $u > 0$, the \acrshort{clpbf} is defined as 
\begin{equation}
%\begin{split}
    \acrshort{clpbf}_{t+1:t+u} = log \left[ \frac{p_{A}(e_{t+u} \mid e_{1:t})}{p_{B}(e_{t+u} \mid e_{1:t})} \right] = \sum_{i=t+1}^{u} log \left[ PL_{i}(A) - PL_{i}(B) \right].
%\end{split} 
\end{equation}
A positive \acrshort{clpbf} indicates evidence in favor for model A and, if $t=0$ and $u=T$, this factor is known as log \acrlong{bf}.
Note that in Section \ref{subsec:hsmm_TuningSMC2}, we mentioned that we usually use $t_0 > 1$ data points to initialize the jitter kernels and then record all future \acrshort{pl} increments going forward, hence the resulting estimate will be the slightly different $\hat{p}(e_{t_{0}+1:T} \mid e_{1:t_0}) = \prod_{t=t_0+1}^T \hat{PL}_{t}$. 
The $t_0$ data points can be seen as training data for the jitter kernels to be initialized in a reasonable parameter region.

\section{Simulation and Experimental Results} \label{sec:hsmm_Experiments}

This section consists of simulation experiments conducted to study the performance of the \acrshort{pf}, the \acrshort{pmcmc} and the \acrshort{smc2} algorithm on \acrlong{hsmm}s. We first generate 1000 data points from a \acrshort{hsmm} with Negative Binomial state duration distributions. A plot with sampled observed and latent data is shown in figure \ref{fig:hsmm_Chp5_HSMM}, which also depicts a \acrshort{pf} estimate of the latent parameter for known continuous model parameter. The advantages of explicitly modelling duration are visible on the third subplot, in which durations in each state vary drastically. The last sub-section shows how the number of hidden states in the data can be estimated using \acrshort{smc2}.

\subsection{Model Dynamics and Prior Assignments}

The model consists of parameter: $\theta = \{\mu, \sigma, p, r, \phi \}$, where the data $e_t \sim N(\mu_{s_t}, \sigma_{s_t})$ has a normal distribution given the latent state. The latent states have the same dynamics as explained in Section \ref{sec:hsmm_HSMM}, latent state and duration  \\ 
    $s_t \sim  \begin{cases} 
        \delta( s_{t}, s_{t-1}) &\text{ $d_{t-1} > 0$ }\\
        Categorical(p_{s_{t-1}}) &\text{ $d_{t-1} = 0$ }
    \end{cases}$, 
    $d_t \sim \begin{cases} 
        \delta( d_{t}, d_{t-1} - 1) &\text{ $d_{t-1} > 0$ }\\
        NegativeBinomial(r_{s_{t}}, \phi_{s_{t}}) &\text{ $d_{t-1} = 0$ }
    \end{cases}$.

The $\mu$ parameter have truncated Normal priors with equal variance and different means, $\mu_1 \sim Normal_{(-100, 0)}(\mu = -2, \sigma = 10^5)$, $\mu_2 \sim Normal_{(0, 100)}(\mu = 2, \sigma = 10^5)$. We assigned a truncated Normal prior for the variances $\sigma \sim Normal_{(0, 10)}(\mu = 2, \sigma = 10^5)$ and for the Negative Binomial parameter, $r \sim Normal_{(0, 100]}(\mu = 10, \sigma = 10^5)$. The second duration distribution parameter, $\phi$, has equal mass from 0 to 1, $\phi \sim Beta(\alpha = 1, \beta = 1)$. Similarly, we assigned a Dirichlet prior for the transition probabilities, $p$, that favors equal weights, $p \sim Dirichlet( \alpha_1 = \alpha_2 = ... = \alpha_k = k),$ where $k =$ number of latent states.

\subsection{Estimation}

If parameter $\theta$ are unknown, both the \acrshort{pmcmc} and \acrshort{smc2} machinery can be used. As the latter builds on the former, we first show estimation results for the \acrlong{pmcmc} run. 
The traceplots of four chains for the continuous model parameter can be seen in figure \ref{fig:hsmm_Chp5_HSMM_PMCMC_CHAIN}, and for the latent state sequence in figure \ref{fig:hsmm_Chp5_HSMM_PMCMC_LATENT}. 
After parameter are initialized from the prior distributions, they rapidly converge to the values used to generate the sample data. 
%Any form of parameter optimization can be applied to obtain an initial estimate, but given convergence occurs fast we did not proceed with this idea. 
Common MCMC output statistics are summarized in Table \ref{table:hsmm_Chp5_HSMM_PMCMC}. 
\acrshort{smc2} results can be seen in figure \ref{fig:hsmm_Chp5_HSMM_SMC_CHAIN} for the model parameter and in figure \ref{fig:hsmm_Chp5_HSMM_SMC_LATENT} for the latent variables. %, and figure \ref{fig:hsmm_Chp5_HSMM_SMC_PREDICT} for predictions.
Just as in the \acrshort{pmcmc} case, samples converge fast toward the typical set.
%, and are discussed in more detail in the corresponding plots referenced above. 
Common MCMC output statistics of the final \acrshort{smc2} iteration for the continuous model parameter are displayed in Table \ref{table:hsmm_Chp5_HSMM_SMC}.

\begin{table}[htp]
\centering
  \begin{tabular}{rrrrrrrrrrrrr}
    \hline\hline
     $\mathbf{\theta}$ & \textbf{True} & \textbf{Mean} & \textbf{MCSE} & \textbf{SD} & \textbf{Rhat} & \textbf{Q2.5} & \textbf{Q25.0} & \textbf{Q50.0} & \textbf{Q75.0} & \textbf{Q97.5} \\\hline      
    $\mu_1$ & -2.0 & -2.03 & 0.01 & 0.39 & 1.00 & -2.50 & -2.18 & -2.0 & -1.85 & -1.55 \\
    $\mu_2$ & 2.0 & 1.78 & 0.00 & 0.22 & 1.00 & 1.59 & 1.72 & 1.78 & 1.84 & 1.96 \\
    $\sigma_1$ & 4.0 & 4.01 & 0.00 & 0.27 & 1.00 & 3.72 & 3.91 & 4.00 & 4.11 & 4.32 \\
    $\sigma_2$ & 2.0 & 2.07 & 0.00 & 0.17  & 1.00 & 1.93 & 2.014 & 2.06 & 2.11 & 2.23 \\
    $r_1$ & 10.0 & 13.64 & 0.01 & 4.06  & 1.00 & 5.46 & 10.64 & 14.01 & 17.09 & 19.72 \\
    $r_2$ & 15.0 & 13.18 & 0.23 & 8.04 & 1.00 & 4.12 & 8.03 & 11.26 & 15.99 & 32.96 \\
    $\phi_1$ & 0.3 & 0.37 & 0.00 & 0.08 & 1.00 & 0.18 & 0.32 & 0.38 & 0.43 & 0.48 \\
    $\phi_2$ & 0.3 & 0.25 & 0.00 & 0.09 & 1.00 & 0.10 & 0.18 & 0.24 & 0.31 & 0.48 \\
    \hline\hline
  \end{tabular}

\caption{Posterior output statistics of 4 PMCMC chains for a HSMM on simulated data in Section \ref{sec:hsmm_Experiments}. 2000 iterations were run with 1000 iterations burnin, resulting in 4000 total samples. Initial parameter have been sampled from the prior distribution.}
\label{table:hsmm_Chp5_HSMM_PMCMC}
\end{table}

\begin{table}[htp]
\centering
  \begin{tabular}{rrrrrrrrrrrrr}
    \hline\hline
     $\mathbf{\theta}$ & \textbf{True} & \textbf{Mean} & \textbf{MCSE} & \textbf{SD} & \textbf{Q2.5} & \textbf{Q25.0} & \textbf{Q50.0} & \textbf{Q75.0} & \textbf{Q97.5} \\\hline      
     $\mu_1$ & -2.0 & -2.04 & 0.01 & 0.42 & -2.86 & -2.28 & -2.05 & -1.82 & -1.06 \\
    $\mu_2$ & 2.0 & 1.7 & 0.00 & 0.17 & 1.42 & 1.63 & 1.73 & 1.83 & 2.10 \\
    $\sigma_1$ & 4.0 & 4.03 & 0.00 & 0.22 & 3.61 & 3.90 & 4.02 & 4.16 & 4.48 \\
    $\sigma_2$ & 2.0 & 2.00 & 0.00 & 0.18 & 1.42 & 1.94 & 2.02 & 2.1 & 2.30 \\
    $r_1$ & 10.0 & 9.38 & 0.19 & 5.99 & 0.09 & 4.94 & 9.68 & 14.19 & 19.40 \\
     $r_2$ & 15.0 & 23.68 & 0.63 & 22.03 & 1.00 & 8.50 & 16.03 & 31.50 & 87.74 \\
    $\phi_1$ & 0.3 & 0.26 & 0.00 & 0.13 & 0.02 & 0.16 & 0.27 & 0.36 & 0.48 \\
    $\phi_2$ & 0.3 & 0.41 & 0.00 & 0.21 & 0.06 & 0.24 & 0.37 & 0.55 & 0.88 \\
    \hline\hline
  \end{tabular}
    \caption{Posterior output statistics for 100 SMC chains on a HSMM at final iterations on simulated data in Section \ref{sec:hsmm_Experiments}.}
    \label{table:hsmm_Chp5_HSMM_SMC}
\end{table}

%%%%%%%%%%%
\subsection{Determining the number of states} \label{subsec:hsmm_ApplicationsStates}

\begin{figure}[htp]
	\centering
	\includegraphics[ width=1.\textwidth ]{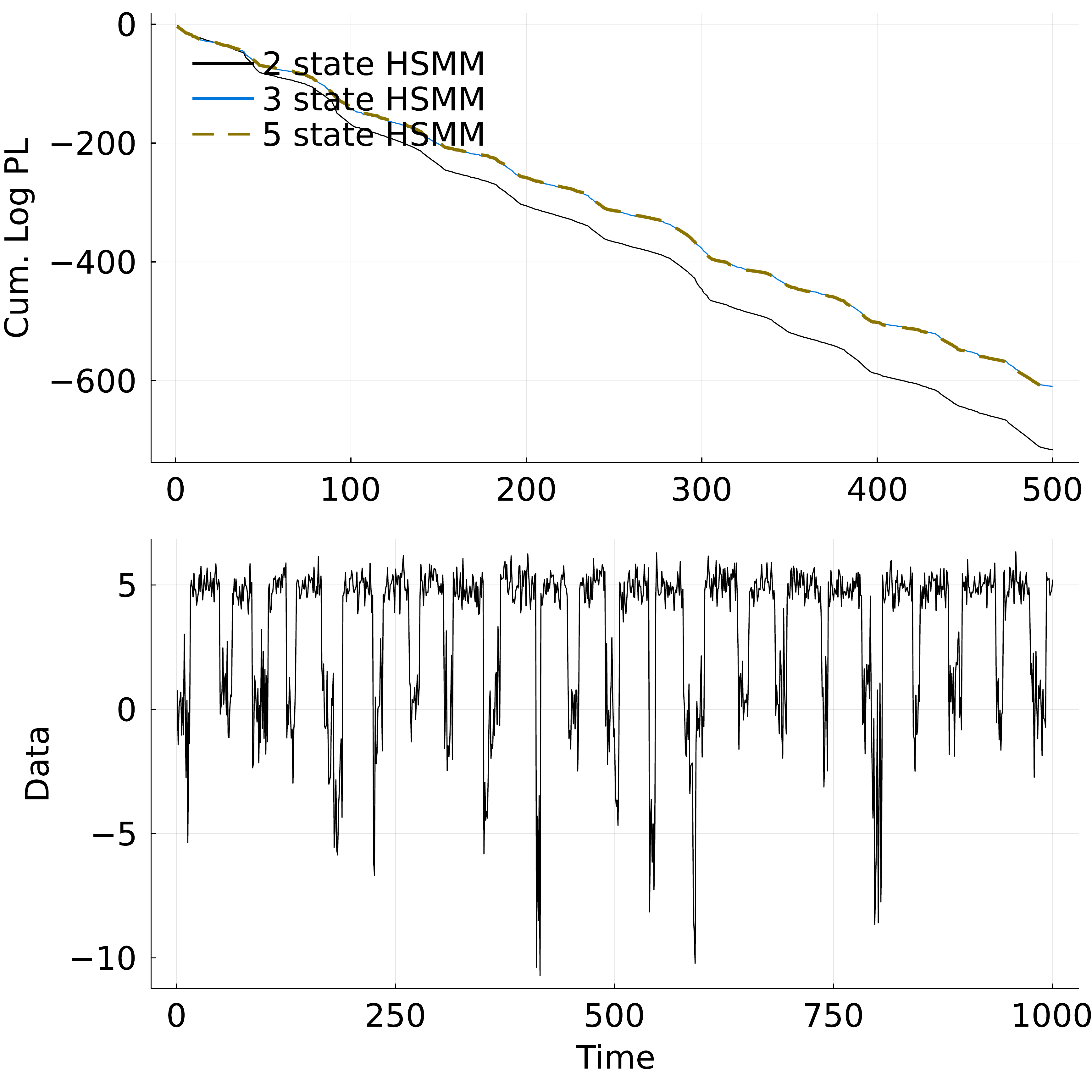}
	\caption{Cumulative log \acrlong{pl} of 2-, 3- and 5-state HSMM, as discussed in Section \ref{sec:hsmm_BayesianInference}. The underlying data has been generated by a 3-state HSMM, and the 5-state HSMM returns \acrshort{pl} levels as the HSMM that generated the data, while the 2-state HSMM is not flexible enough to detect the data structure. Hence, we can "overfit" the number of states and let the algorithm decide the number of states itself in case no prior knowledge is available. The bottom plot shows the whole data sequence, while the bottom graph depicts the cumulative log \acrshort{pl} as defined in Section \ref{subsec:hsmm_ModelSelection} of the last 500 iterations.}
	\label{fig:hsmm_Chp6_DetermineStates_MarginalLik}
\end{figure}

Before estimating model parameter for a given data set, the researcher has to choose the number of hidden states in a discrete \acrshort{ssm}, which is often challenging a priori. To tackle this problem, we choose methods from the overfitting mixtures literature, see \citet{rousseau11, frue06, frue18}. In particular, \citet{rousseau11} show that the posterior distribution of a mixture model is much more stable if the prior weights of the mixture components are concentrated on the boundary regions of the parameter space. We borrow this concept for \acrshort{ssm}s and show that by assigning appropriate prior weights on the transition matrices of a \acrshort{hsmm}, we can choose more states than necessarily describe the data and still have an interpretable structure, as superfluous latent states will never be visited during the estimation process. As an experiment, we fit a 2-, 3-, and 5-state HSMM to data that was generated by a 3-state HSMM. The parameter used to generate the data can be seen in Table \ref{tab:Chp6_DetermineStates_parameter}, and the corresponding data in plot \ref{fig:hsmm_Chp6_DetermineStates_MarginalLik}. 
In Section \ref{sec:hsmm_BayesianInference}, we mentioned that the \acrlong{pl} can be estimated as a by-product of the inference procedure in \acrshort{smc2}. Hence, we tracked the cumulative log \acrshort{pl} for each model at each iteration, and plotted the results in figure \ref{fig:hsmm_Chp6_DetermineStates_MarginalLik}. It can be seen that the 3- and 5-state \acrshort{hsmm}s \acrlong{pl} perfectly align, and we conclude that it is possible to determine the number of states via \acrshort{smc2}, and even check for differences during the data propagation. 
All estimates for the model parameter can be seen in Table \ref{tab:Chp6_DetermineStates_parameter}. In this table, state $2$ and $4$ are almost never visited and act as superfluous regimes. The remaining state estimates contain the correct parameter in the 95\% \acrlong{ci}s.

\begin{table}[htp]
\centering
   \begin{tabular}{rrrrrrrrrrrrr}
    \hline\hline
    $\mathbf{\theta}$ & \textbf{True} & \textbf{Mean} & \textbf{MCSE} & \textbf{SD} & \textbf{Q2.5} & \textbf{Q25.0} & \textbf{Q50.0} & \textbf{Q75.0} & \textbf{Q97.5} \\\hline
    $\mu_1$ & -5.0 & -4.73 & 0.03 & 1.48 & -8.21 & -5.43 & -4.66 & -3.97 & -1.6 \\
    $\mu_2$ & - & -1.5 & 0.1 & 4.18 & -8.8 & -4.46 & -1.04 & 0.09 & 8.64 \\
    $\mu_3$ & 0.0 & -0.44 & 0.07 & 2.06 & -6.06 & -0.19 & -0.05 & 0.07 & 2.93 \\
    $\mu_4$ & - & -1.56 & 0.11 & 5.36 & -9.25 & -5.65 & -3.19 & 2.56 & 9.35 \\
    $\mu_5$ & 5.0 & 4.95 & 0.0 & 0.03 & 4.9 & 4.94 & 4.96 & 4.97 & 5.01 \\
    $\sigma_1$ & 2.5 & 2.95 & 0.03 & 1.38 & 1.13 & 2.26 & 2.7 & 3.2 & 7.69 \\
    $\sigma_2$ & - & 3.27 & 0.08 & 2.32 & 0.76 & 1.56 & 2.42 & 4.11 & 9.24 \\
    $\sigma_3$ & 1.5 & 1.92 & 0.05 & 1.33 & 1.27 & 1.46 & 1.55 & 1.67 & 6.9 \\
    $\sigma_4$ & - & 4.43 & 0.06 & 2.68 & 0.46 & 2.4 & 3.59 & 6.68 & 9.62 \\
    $\sigma_5$ & 0.5 & 0.52 & 0.0 & 0.06 & 0.45 & 0.49 & 0.51 & 0.53 & 0.69 \\
    $\lambda_1$ & 5.0 & 4.53 & 0.04 & 1.7 & 1.38 & 3.4 & 4.46 & 5.53 & 8.36 \\
    $\lambda_2$ & - & 10.14 & 0.17 & 6.71 & 1.55 & 5.08 & 9.33 & 12.23 & 27.51 \\
    $\lambda_3$ & 10.0 & 10.76 & 0.1 & 3.58 & 3.16 & 9.89 & 10.8 & 11.65 & 21.89 \\
    $\lambda_4$ & - & 11.72 & 0.23 & 8.77 & 1.36 & 4.01 & 8.9 & 18.78 & 29.02 \\
    $\lambda_5$ & 30.0 & 30.92 & 0.04 & 1.59 & 27.72 & 29.84 & 30.98 & 32.03 & 33.89 \\
    $p_{1,1}$ & - \\
    $p_{1,2}$ & - & 0.15 & 0.01 & 0.22 & 0.0 & 0.0 & 0.05 & 0.26 & 0.77 \\
    $p_{1,3}$ & 0.2 & 0.27 & 0.01 & 0.26 & 0.0 & 0.05 & 0.25 & 0.46 & 0.87 \\
    $p_{1,4}$ & - & 0.08 & 0.0 & 0.18 & 0.0 & 0.0 & 0.01 & 0.09 & 0.74 \\
    $p_{1,5}$ & 0.8 & 0.50 \\

    $p_{2,1}$ & - & 0.19 & 0.01 & 0.25 & 0.0 & 0.01 & 0.07 & 0.27 & 0.9 \\
    $p_{2,2}$ & - \\
    $p_{2,3}$ & - & 0.2 & 0.01 & 0.26 & 0.0 & 0.0 & 0.07 & 0.31 & 0.92 \\
    $p_{2,4}$ & - & 0.13 & 0.01 & 0.23 & 0.0 & 0.0 & 0.02 & 0.14 & 0.89 \\
    $p_{2,5}$ & - \\

    $p_{3,1}$ & 0.2 & 0.17 & 0.0 & 0.17 & 0.0 & 0.03 & 0.13 & 0.26 & 0.59 \\
    $p_{3,2}$ & - & 0.1 & 0.01 & 0.16 & 0.0 & 0.0 & 0.03 & 0.14 & 0.59 \\
    $p_{3,3}$ & - \\
    $p_{3,4}$ & - & 0.07 & 0.0 & 0.14 & 0.0 & 0.0 & 0.01 & 0.07 & 0.49 \\
    $p_{3,5}$ & 0.8 & 0.67 \\

    $p_{4,1}$ & - & 0.23 & 0.01 & 0.29 & 0.0 & 0.01 & 0.08 & 0.37 & 0.95 \\
    $p_{4,2}$ & - & 0.22 & 0.01 & 0.29 & 0.0 & 0.0 & 0.07 & 0.36 & 0.94 \\
    $p_{4,3}$ & - & 0.24 & 0.01 & 0.29 & 0.0 & 0.01 & 0.11 & 0.4 & 0.94 \\
    $p_{4,4}$ & - \\
    $p_{4,5}$ & - \\

    $p_{5,1}$ & 0.2 & 0.2 & 0.0 & 0.15 & 0.0 & 0.07 & 0.19 & 0.3 & 0.51 \\
    $p_{5,2}$ & - & 0.18 & 0.01 & 0.26 & 0.0 & 0.0 & 0.04 & 0.24 & 0.82 \\
    $p_{5,3}$ & 0.8 & 0.57 & 0.01 & 0.28 & 0.0 & 0.47 & 0.67 & 0.78 & 0.89 \\
    $p_{5,4}$ & - \\
    $p_{5,5}$ & - \\
    \hline\hline
  \end{tabular}
  \caption{Parameter estimates for 5-state \acrshort{hsmm}. The first column depicts the parameter for the 3 state \acrshort{hsmm} that were used to generate the data. State 2 and 4 are hardly visited, and are the superfluous states. $p_{i,j}$ denotes the transition probability from state $i$ to $j$. $p_{i,i}$ is separately modeled by the duration distribution. The transition probabilities $p$ form a simplex, so the probability to transition to the final state is determined by all other states.}
  \label{tab:Chp6_DetermineStates_parameter}
\end{table}

%%%%%%%%%%%%%%%%%%%%%%%%%%%
%Chapter - Comments

\section{Applications on VIX Times Series Data} \label{sec:hsmm_Applications}

This section consists of two interesting applications for \acrlong{ssm}s. 
The first contains a case study on model selection and prediction for a financial data set, the other displays how \acrshort{ssm}s can be used for clustering purposes.

%%%%%%%%%%%
\subsection{Prediction and Model Selection on Financial Volatility} \label{sec:ApplicationsVIX}

%Financial data is notoriously difficult to predict due to their sequential structure. 
In this Section, we are modeling VIX index data directly via \acrshort{hsmm}s and other popular benchmark models. 
The VIX index is derived from options with near-term expiration dates on the US major equity S\&P~500 index, and is a popular indicator for future short term volatility expectations.
Volatility is often modeled indirectly based on (log) stock prices via \acrfull{sv} models and its autoregressive nature is well recognized, see e.g. \citep{hull:1987, kim:1998, Kastner16}. We incorporate this behaviour in our model by assigning autoregressive weights to the location parameters in each latent regime. 
The corresponding model can be viewed as a separate \acrshort{ar}(1) model in each state, where a latent variable following a semi-Markov chain governs the regime changes and durations.
Multiple duration distribution choices, such as the negative Binomial and Poisson distribution, are tested and benchmarked  against \acrshort{hmm} and AR(1) models.

For real data collection, we first obtain 1000 daily end-of-day data points of this instrument, which can be seen in figure \ref{fig:hsmm_Chp6_realdata}, as of January 1st 2022 from the Thomson Reuters database.
Notably, the Covid-19 epidemic is included in the data, which causes the evidence to be much more volatile from the start of 2020 and to transition to a different regime from there onward.
%We are fitting autoregressive \acrshort{hsmm}s with negative Binomial as well as Poisson distribution to this data, and compare it against standard \acrshort{hmm} and AR(1) models.  
Common ideas to address this issue are to assign change points across the times series before estimating the model parameter. Such methods have an easily interpretable structure, but lack the 
information from a stochastic process governing the model dynamics, which enhances inference capabilities for the data.
For example, due to the discrete state space formulation of our proposed model, parameter interpretation is straightforward, while 
the \acrshort{hsmm} dynamics allow to incorporate significantly more decision-making tools, such as regime change and duration forecasting.

\subsubsection{Model Dynamics and Prior Assignments}

The AR(1) HSMM with Negative Binomial duration consists of parameter: $\theta = \{\mu, \sigma, w, p, r, \phi \}$, where the data $e_t \sim N(w_{s_t} \times e_{t-1} + \mu_{s_t}, \sigma_{s_t})$ has a normal distribution given the latent state and the previous data point. 

The latent states have the same dynamics as explained in Section \ref{sec:hsmm_HSMM}, latent state and duration \\
    $s_t \sim  \begin{cases} 
        \delta( s_{t}, s_{t-1}) &\text{ $d_{t-1} > 0$ }\\
        Categorical(p_{s_{t-1}}) &\text{ $d_{t-1} = 0$ }
    \end{cases}$,
    $d_t \sim \begin{cases} 
        \delta( d_{t}, d_{t-1} - 1) &\text{ $d_{t-1} > 0$ }\\
        NegativeBinomial(r_{s_{t}}, \phi_{s_{t}}) &\text{ $d_{t-1} = 0$ }
    \end{cases}$.

The $\mu$ parameter have truncated Normal priors with equal variance and different means, $\mu_1 \sim Normal_{(0, 10)}(\mu = 3, \sigma = 10^5)$, $\mu_2 \sim Normal_{(0, 10)}(\mu = 1, \sigma = 10^5)$. We assigned a truncated Normal prior for the variances $\sigma_{1,2} \sim Normal_{(0, 10)}(\mu = 0.2, \sigma = 10^5)$ and for the Negative Binomial parameter, $r_1 \sim Normal_{(0, 100]}(\mu = 5, \sigma = 10^5)$, $r_2 \sim Normal_{(0, 100]}(\mu = 2.5, \sigma = 10^5)$. The second duration distribution parameter $\phi$ has equal mass from 0 to 1, $\phi_{1,2} \sim Beta(\alpha = 1, \beta = 1)$. Similarly, we assigned a Dirichlet prior for the transition probabilities $p$ that favors equal weights, $p \sim Dirichlet( \alpha_1 = \alpha_2 = ... = \alpha_k = k),$ where $k =$ number of latent states. The autoregressive parameter, $w$, is bounded between $-1$ and $1$ by assigning a truncated prior, $w_{1,2} \sim Normal_{(-1, 1)}(\mu = 0, \sigma = 10^5)$. The intuition behind the prior assignment is that there is a single state with a large constant that mimics sudden jumps in the index, while the other state has a much smaller constant, but possibly a higher autoregressive weight $w$. This assumption is confirmed in the actual estimates, visible in figure \ref{fig:hsmm_Chp6_SMC2Chain}.
The AR(1) HSMM with Poisson duration only differentiates with respect to the duration parameter, $\theta = \{\mu, \sigma, w, p, \lambda \}$. In this case, the latent duration has a Poisson distribution, \\
$d_t \sim \begin{cases}
            \delta( d_{t}, d_{t-1} - 1) &\text{ $d_{t-1} > 0$ }\\
            Poisson(\lambda_{s_{t}}) &\text{ $d_{t-1} = 0$ }
\end{cases}$, 
where $\lambda \sim Normal_{(0, 100]}(\mu = 20, \sigma = 10^5)$.
The AR(1) HMM with parameter $\theta = \{\mu, \sigma, w, p\}$, has the same data dynamics and latent state dynamics $s_t \sim Categorical(p_{s_{t-1}})$. All prior configurations are assigned from the previous models.
The AR(1) Model with parameter $\theta = \{\mu, \sigma, w \}$ does not have latent variables, and data dynamics $e_t \sim Normal(w \times e_{t-1} + \mu, \sigma)$. The priors for $\mu$ are set to $\mu \sim Normal_{(0, 10)}(\mu = 2.0, \sigma = 10^5)$, where the location parameter has been set as the mean of the location parameter for the 2 states in the state space models defined above. $\sigma$ and $w$ have the same prior as the other models.

\subsubsection{Results}
\begin{figure}[htp]
	\centering
	\includegraphics[ width=1.\textwidth ]{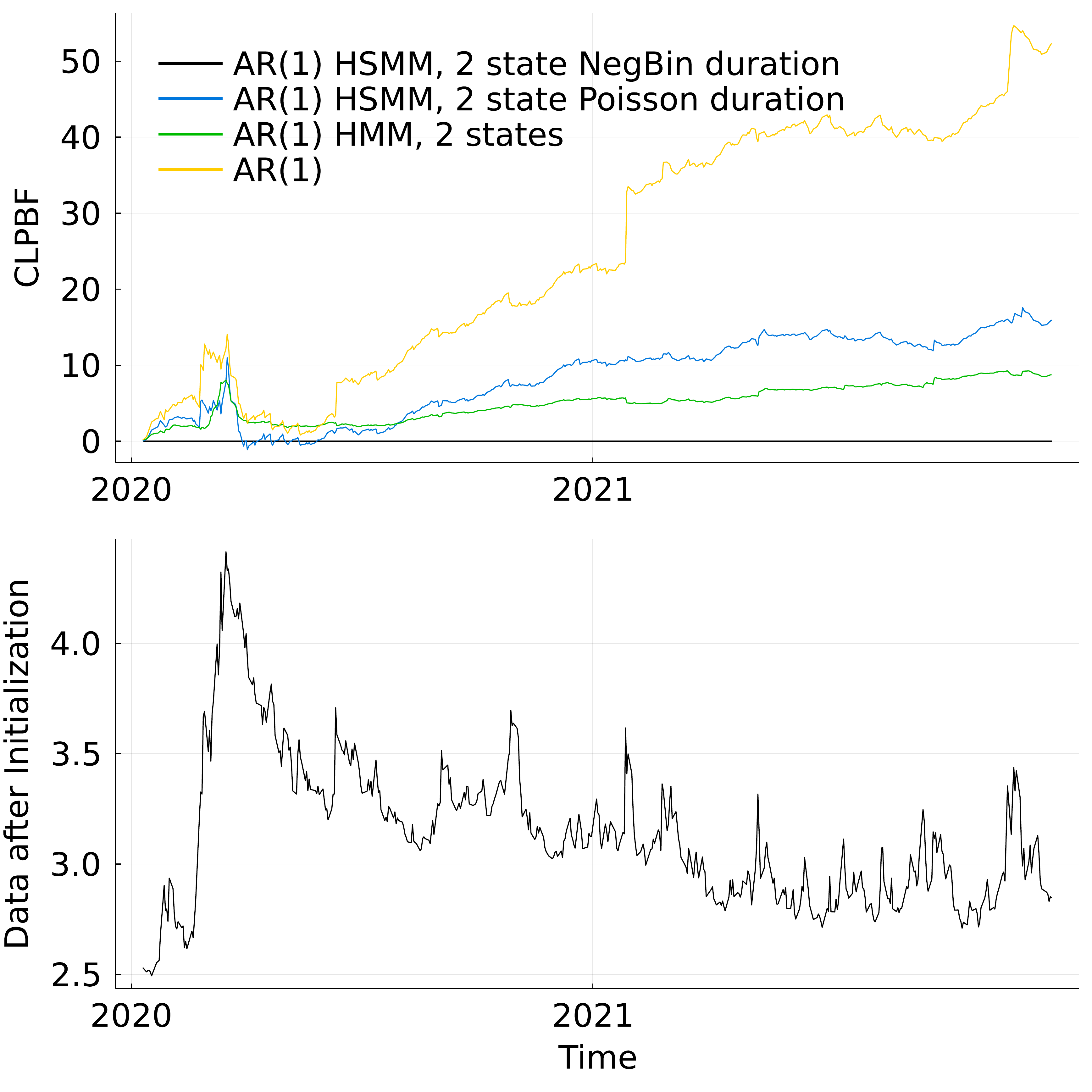}
	\caption{
        The top plot depicts the \acrlong{clpbf} as defined in Section \ref{subsec:hsmm_ModelSelection} of the winning model in Section \ref{sec:hsmm_Applications} at each iteration. At the bottom, the corresponding log VIX index data is shown over time.
        }
	\label{fig:hsmm_Chp6_BayesFactor}
\end{figure}

A detailed methodology for model comparison can be found in section \ref{subsec:hsmm_ModelSelection}. The \acrshort{smc2} machinery with the tuning configurations discussed in Section \ref{sec:hsmm_BayesianInference} is used to estimate model parameter. Results for our proposed model can be seen in Figure \ref{fig:hsmm_Chp6_SMC2Chain} and Table \ref{table:hsmm_Chp6_HSMM_SMC} for model parameter. The traceplots show all model parameter estimates at each point in time with the corresponding 95\% credible interval. 
There is a clear distinction between a lower and higher volatility state, in which the latter has volatility levels of roughly four times the normal state. 
Interestingly, there seems to be a regime shift at the time COVID-19 makes significant headlines in the global markets, which is captured very fast during the estimation process, and would be impossible for standard batch estimation methods. 
At the start of March 2020, the volatility parameter in the high-volatility state changes from $0.15$ to $0.20$ within days, and continues to be around this level until the end of the times series, making the regime change easily recognizable.
Figure \ref{fig:hsmm_Chp6_SMC2Latent} 
shows the filtered state trajectory of the latent variables at each time step. 
A re-scaled posterior mean of the latent variable at each time index is shown against the real data in the bottom sub-plot, which displays the clear distinction between a volatile and a stable state in the times series. There is more variation at the initial stages before more data is added to the algorithm. Figure \ref{fig:hsmm_Chp6_SMC2Prediction} depicts model predictions against the realized data at each time step. 
The \acrlong{clpbf} as discussed in Section \ref{sec:hsmm_BayesianInference} is shown in figure \ref{fig:hsmm_Chp6_BayesFactor}. 
The top plot in this figure shows the \acrshort{clpbf} of the AR(1) \acrshort{hsmm} with negative Binomial duration distribution against the other models, which is positive and in favor of the proposed model consistently over time against all other models. 
Diagnostics at the final time period are summarized in Table \ref{table:hsmm_Chp6_pl}.

\begin{table}[htp]
\centering
  \begin{tabular}{rrrrrrrrr}
    \hline\hline
     $\mathbf{\theta}$ & \textbf{Mean} & \textbf{MCSE} & \textbf{SD} & \textbf{Q2.5} & \textbf{Q25.0} & \textbf{Q50.0} & \textbf{Q75.0} & \textbf{Q97.5} \\\hline
    $\mu_1$ & 1.03 & 0.00 & 0.06 & 0.92 & 0.99 & 1.02 & 1.06 & 1.18 \\
    $\mu_2$ & 0.11 & 0.00 & 0.06 & 0.038 & 0.07 & 0.08 & 0.11 & 0.31 \\
    $\sigma_1$ & 0.19 & 0.00 & 0.02 & 0.12 & 0.17 & 0.19 & 0.21 & 0.24 \\
    $\sigma_2$ & 0.06 & 0.0 & 0.00 & 0.04 & 0.06 & 0.05 & 0.06 & 0.06 \\
    $w_1$ & 0.68 & 0.0 & 0.01 & 0.65 & 0.68 & 0.69 & 0.69 & 0.7 \\
    $w_2$ & 0.96 & 0.00 & 0.03 & 0.88 & 0.96 & 0.97 & 0.97 & 0.98 \\
    $r_1$ & 8.39 & 0.38 & 11.57 & 0.15 & 0.64 & 2.82 & 11.01 & 42.44 \\
    $r_2$ & 0.41 & 0.03 & 0.44 & 0.14 & 0.24 & 0.33 & 0.46 & 1.01 \\
    $\phi_1$ & 0.64 & 0.0 & 0.28 & 0.11 & 0.38 & 0.71 & 0.90 & 0.98 \\
    $\phi_2$ & 0.03 & 0.00 & 0.02 & 0.01 & 0.02 & 0.02 & 0.03 & 0.06 \\
    \hline\hline
  \end{tabular}
  \caption{Posterior output statistics for 100 SMC chains on HSMM in chapter \ref{sec:hsmm_Applications} at final iterations in \acrshort{smc2} run.}
\label{table:hsmm_Chp6_HSMM_SMC}
\end{table}

\begin{table}[htp]
\centering
  \begin{tabular}{rrr}
    \hline\hline
    \textbf{Names} & \textbf{ Cum. Log PL} & \textbf{Diff. to $\mathbf{(\star)}$} \\\hline
    $(\star)$ AR(1) HSMM - 2 states - Neg. Bin. Duration & 550.85 & 0.00 \\
    AR(1) HSMM - 2 states - Poisson Duration & 534.90 & -15.94 \\
    AR(1) HMM - 2 states & 542.1 & -8.75 \\
    AR(1) Model & 498.51 & -52.33 \\
    \hline\hline
  \end{tabular}
  \caption{Cumulative log \acrlong{pl} as defined in Section \ref{subsec:hsmm_ModelSelection} for various discrete state space models fitted via \acrshort{smc2} on data described in chapter \ref{sec:hsmm_Applications}.}
  \label{table:hsmm_Chp6_pl}
\end{table}

%%%%%%%%%%%
\subsection{Clustering} 
\label{sec:Clustering}

\begin{figure}[htp]
	\centering
	\includegraphics[ width=1.\textwidth ]{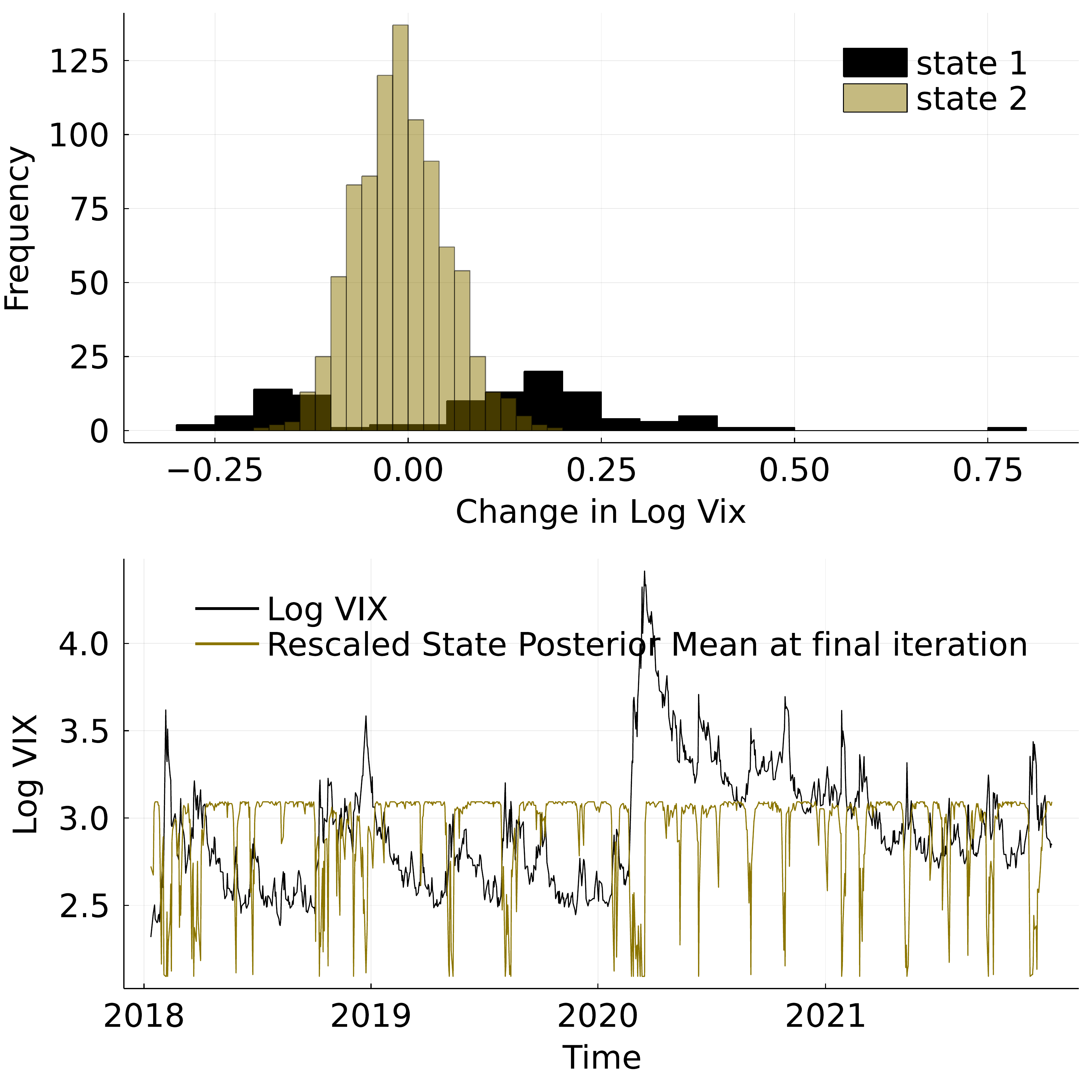}
	\caption{The top plot depicts a histogram for changes in log VIX data, conditioned on the most probable posterior latent state from the final \acrshort{smc2} iteration. The bottom graph displays the log VIX data over time (black) alongside a re-scaled state posterior mean from the final \acrshort{smc2} iteration (gold).}
	\label{fig:hsmm_Chp6_ClusterSummary}
\end{figure}

Once parameter have been estimated, the latent state trajectory estimates can be used to cluster the log VIX data into different regimes. A re-scaled state posterior mean of the latent trajectory at the last \acrshort{smc2} against the actual data can be seen at the top plot in Figure \ref{fig:hsmm_Chp6_ClusterSummary}. 
State 1 corresponds to a short duration state with jumps and drastic changes in levels, 
while state 2 corresponds to a regime with more normal volatility levels that is observed for the majority of times.
Based on the posterior mean, we clustered the changes in the index for each state, which can be seen in the bottom plot of Figure \ref{fig:hsmm_Chp6_ClusterSummary}. Here, state 1 is clearly associated with drastic movements in either direction, while state 2 represents a more normal market environment.  

%%%%%%%%%%%%%%%%%%%%%%%%%%%
%Chapter - Comments

\section{Conclusions} \label{sec:hsmm_Conclusion}
%Overview
In this paper, we discussed sequential parameter estimation techniques for \acrlong{ssm}s with a focus on \acrlong{hsmm}s. % and propose various tuning mechanisms in chapter \ref{sec:hsmm_Experiments}. 
We compared the forecasting accuracy of various models on financial data and concluded that the \acrshort{hsmm} has a superior predictive performance against other popular discrete \acrshort{ssm}s. 
Moreover, we demonstrated how by-products emerging from a \acrshort{smc2} estimation run can be used to determine the number of latent regimes governing such models. 
While the additional duration variable in the \acrshort{hsmm} typically causes parameter inference to be more challenging due to the increased computational costs of the likelihood function, it adds significantly more flexibility in modelling the latent process. 
Our proposed inference technique has the same computational costs for both the basic \acrshort{hmm} and the \acrshort{hsmm} and is particularly suitable for sequential data. % Open research questions
%As for future research areas, we observed that the likelihood estimate variance for a particle filter is not constant across the whole parameter range for a variety of \acrshort{ssm}s. 
As for future research topics, more optimized techniques to adaptively select the number of particles in a \acrshort{pf} may lead to faster runs and improved mixing for both the \acrshort{pmcmc} and \acrshort{smc2} algorithm. 
Furthermore, as tuning the individual \acrshort{mcmc} and \acrshort{pf} kernels was handled independently during the \acrshort{smc2} runs, adding information from all chains may drastically increase the tuning process for the individual jitter kernels.
%Furthermore, in order to initialize the jitter kernels in the \acrshort{smc2} algorithm in a reasonable parameter region, $t_0 > 1$ initial data points have been used when fitting the data. Finding a good value for $t_0$ can drastically increase the mixing of the different chains and likewise decrease the time spent in the \acrshort{smc2} run. 

%Open research questions include finding the minimum number of initial data points for the initialization of the \acrshort{smc2} algorithm. Few enough that initial computation is cheap and covers a broad range of values for each parameter, such that the algorithm does not depend on initial parameter, but enough data points that a difference in regimes may be recognized. This will, of course, depend on the number of transitions to each state, which is unknown a priori. 

%%%%%%%%%%%%%%%%%%%%%%%%%%%
%Chapter - Comments
\section{Software} \label{sec:hsmm_Sofware}

The data and code used to run the algorithms in this paper can be be seen in \url{https://github.com/paschermayr/Publish_SequentialHSMM}. For more detailed information about the implementations for running all algorithms and computing all tables can be found in \url{https://github.com/paschermayr/Baytes.jl} and its sub-libraries. The corresponding plots are defined in \url{https://github.com/paschermayr/BaytesInference.jl}.

%%%%%%%%%%%%%%%%%%%%%%%%%%%
%Chapter - Comments
\clearpage
\bibliography{_references}

%%%%%%%%%%%%%%%%%%%%%%%%%%%
%Chapter - Comments
\clearpage
\printglossary[type=\acronymtype]

%%%%%%%%%%%%%%%%%%%%%%%%%%%
%Chapter - Comments
\clearpage
\clearpage
\appendix

%%%%%%%%%%%%%%%%%%%%%%%%%%%%%%%%%%%%%%%%%%%%%%%%%%%%%%%%%%%%%%%%%%%%%%%%% 
%%%%%%%%%%%%%%%%%%%%%%%%%%%%%%%%%%%%%%%%%%%%%%%%%%%%%%%%%%%%%%%%%%%%%%%%%
%%%%%%%%%%%%%%%%%%%%%%%%%%%%%%%%%%%%%%%%%%%%%%%%%%%%%%%%%%%%%%%%%%%%%%%%%
%\section{Appendix} \label{sec:App_Appendix}

%%%%%%%%%%%%%%%%%%%%%%%%%%%
%Pseudo Algorithms
%\clearpage
%\input{other/Appendix_Models}

%%%%%%%%%%%%%%%%%%%%%%%%%%%
%Pseudo Algorithms
\clearpage
%%%%%%%%%%%%%%%%%%%%%%%%%%%%%%%%%%%%%%%%%%%%%%%%%%%%%%%%%%%%%%%%%%%%%%%%%
%%%%%%%%%%%%%%%%%%%%%%%%%%%%%%%%%%%%%%%%%%%%%%%%%%%%%%%%%%%%%%%%%%%%%%%%%
%%%%%%%%%%%%%%%%%%%%%%%%%%%%%%%%%%%%%%%%%%%%%%%%%%%%%%%%%%%%%%%%%%%%%%%%%
\section{Plots} \label{sec:App_Plots}

% Chp2 - HSMM
%%%%%%%%%%%%%%%%%%%%%%%%%%%%%%%%%%%%%%%%%%%%%%%%%%%%%%%%%%%%%%%%%%%%%%%%%
% Bayesian HSMM
%\begin{figure}[htp]
 %   \centering
%	\includegraphics[ width=1.\textwidth ]{_figs/Chp2_BayesianHSMM_NoParam.pdf}
%	\caption{$K$-state HSMM. The upper random variable $d$ denotes the remaining time spent in any particular hidden state $s$. $e$ denotes the observed observation at each time point.}
 %   \label{fig:hsmm_Chp3_BayesianHSMM_NoParam}
%\end{figure}

% Chp3 - Inference
%%%%%%%%%%%%%%%%%%%%%%%%%%%%%%%%%%%%%%%%%%%%%%%%%%%%%%%%%%%%%%%%%%%%%%%%%
% PF Tuning - pick N

\begin{figure}[htp]
	\centering
	\includegraphics[ width=1.\textwidth ]{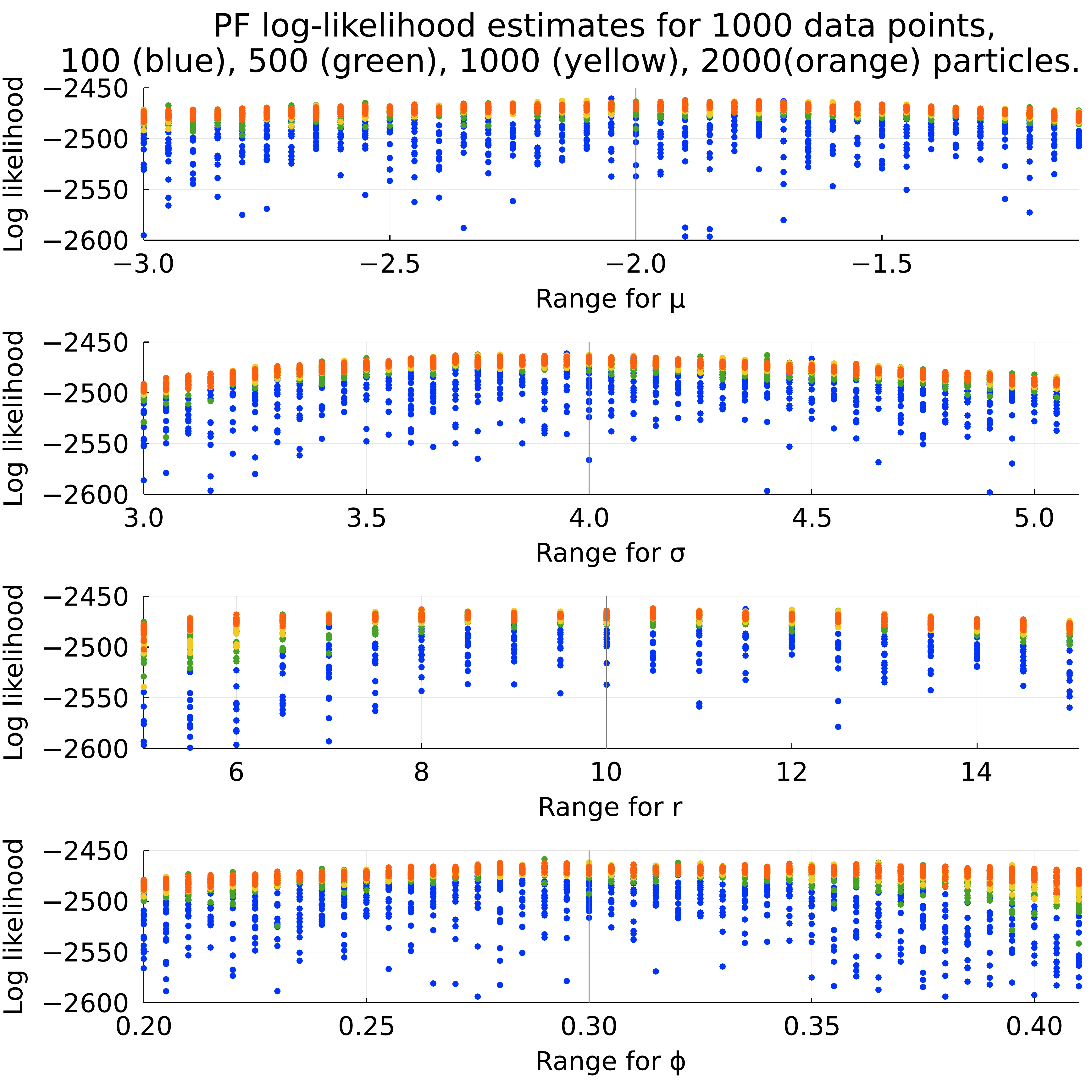}
	\caption{This graph shows \acrlong{pf} likelihood estimates for a range of different parameter values. At each column, all parameter were kept constant except the labeled parameter at the x-axis. The different colors depict various amount of particles used for the computations: 100 (blue), 500 (green), 1000 (yellow), 2000 (orange) particles for sample data of size 1000.}
	\label{fig:hsmm_Chp3_HSMM_ParticleFilterEstimate}
\end{figure}

\begin{figure}[htp]
	\centering
	\includegraphics[ width=1.\textwidth ]{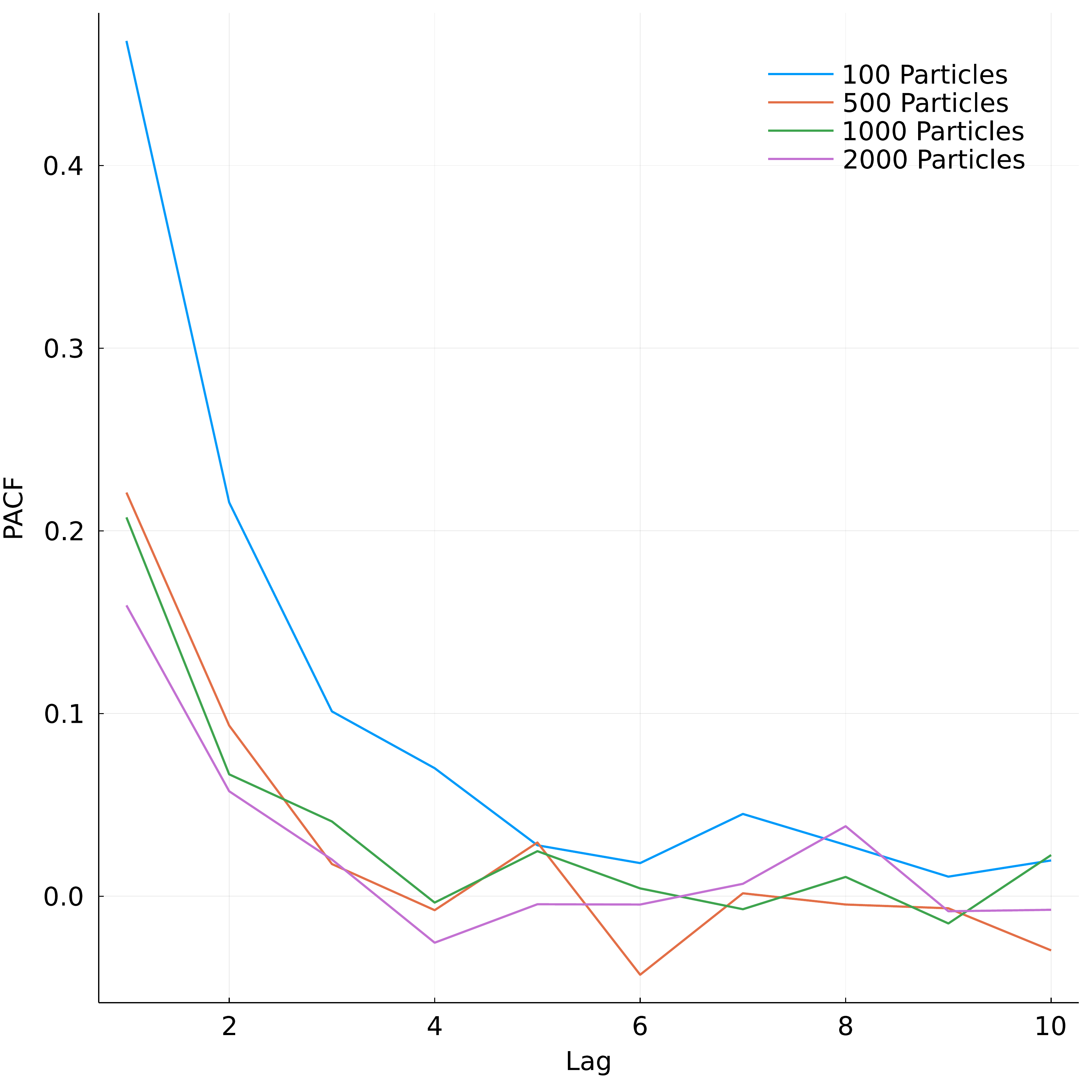}
	\caption{This graph shows the \acrlong{pacf} of \acrlong{pf} likelihood estimates that have been obtained during 1000 \acrshort{pmcmc} steps after burnin. The \acrlong{pf} in the \acrshort{pmcmc} kernels were set to have a different amount of particles for each run.}
	\label{fig:hsmm_Chp3_HSMM_PACF}
\end{figure}

% Chp4 - Simulation
%%%%%%%%%%%%%%%%%%%%%%%%%%%%%%%%%%%%%%%%%%%%%%%%%%%%%%%%%%%%%%%%%%%%%%%%%
% HSMM

%%%%%%%%%%%%%%%%%%%%%%%%%%%%%%%%%%%%%%%%%%%%%%%%%%%%%%%%%%%%%%%%%%%%%%%%%
% Simulated Data

\begin{figure}[htp]
	\centering
	\includegraphics[ width=1.\textwidth ]{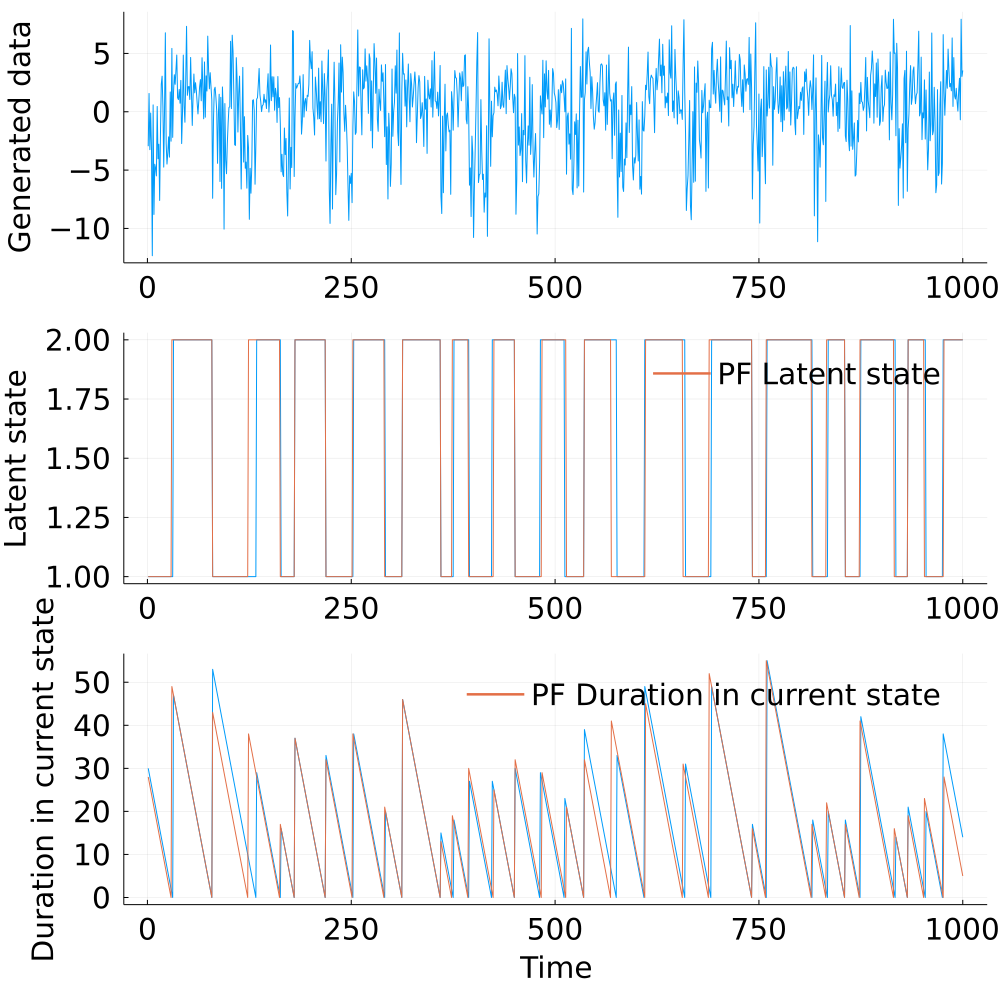}
	\caption{
        The upper graph shows generated observed data by the \acrshort{hsmm} depicted in section \ref{sec:hsmm_Experiments}. The middle plot shows the hidden state (blue) and a sample of a filtered trajectory from a particle filter. The lower plot shows the remaining duration given the current state, and a sample of a filtered particles from a particle filter.}
	\label{fig:hsmm_Chp5_HSMM}
\end{figure}

%%%%%%%%%%%%%%%%%%%%%%%%%%%%%%%%%%%%%%%%%%%%%%%%%%%%%%%%%%%%%%%%%%%%%%%%%
% PMCMC
\begin{figure}[htp]
	\centering
	\includegraphics[ width=1.\textwidth ]{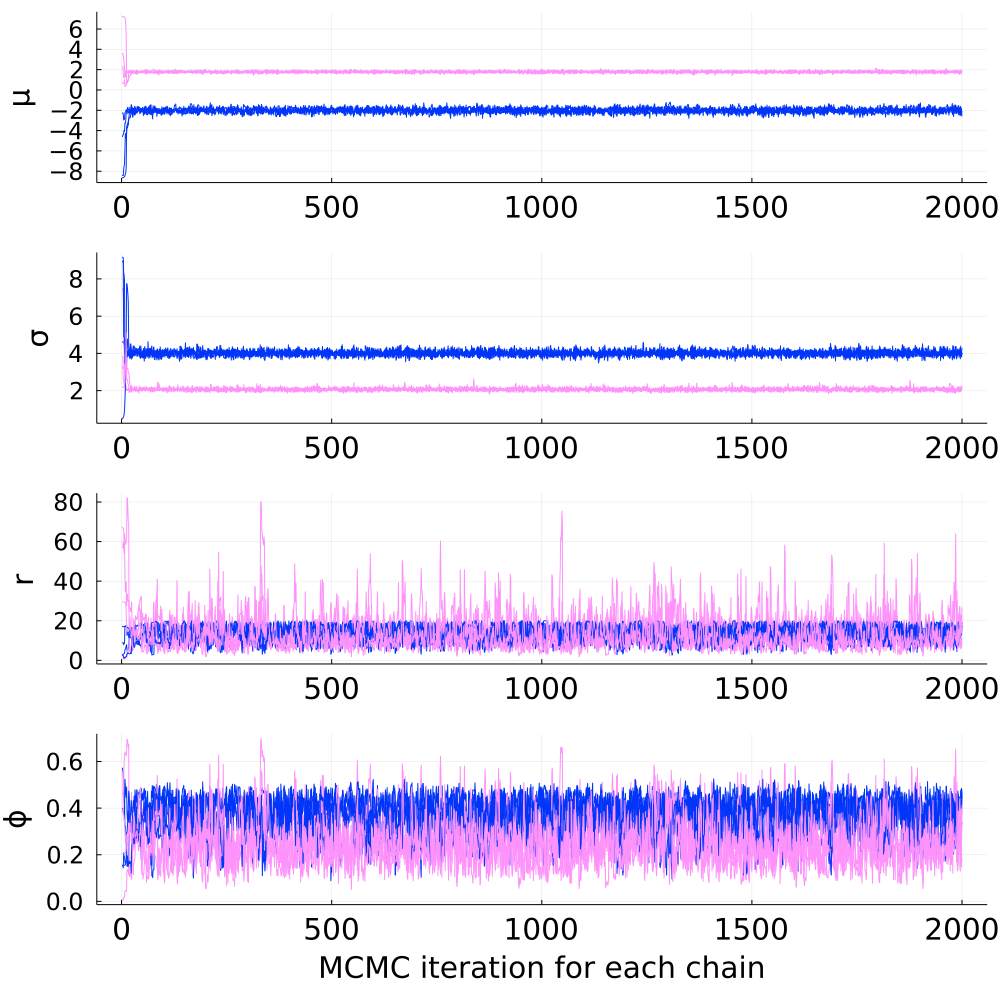}
	\caption{Traceplots of four \acrlong{pmcmc} chains for continuous model parameter of a \acrshort{hsmm} in section \ref{sec:hsmm_Experiments}. %Parameter used to generate sample data are shown as dashed horizontal lines.
 }
	\label{fig:hsmm_Chp5_HSMM_PMCMC_CHAIN}
\end{figure}
\begin{figure}[htp]
	\centering
	\includegraphics[ width=1.\textwidth ]{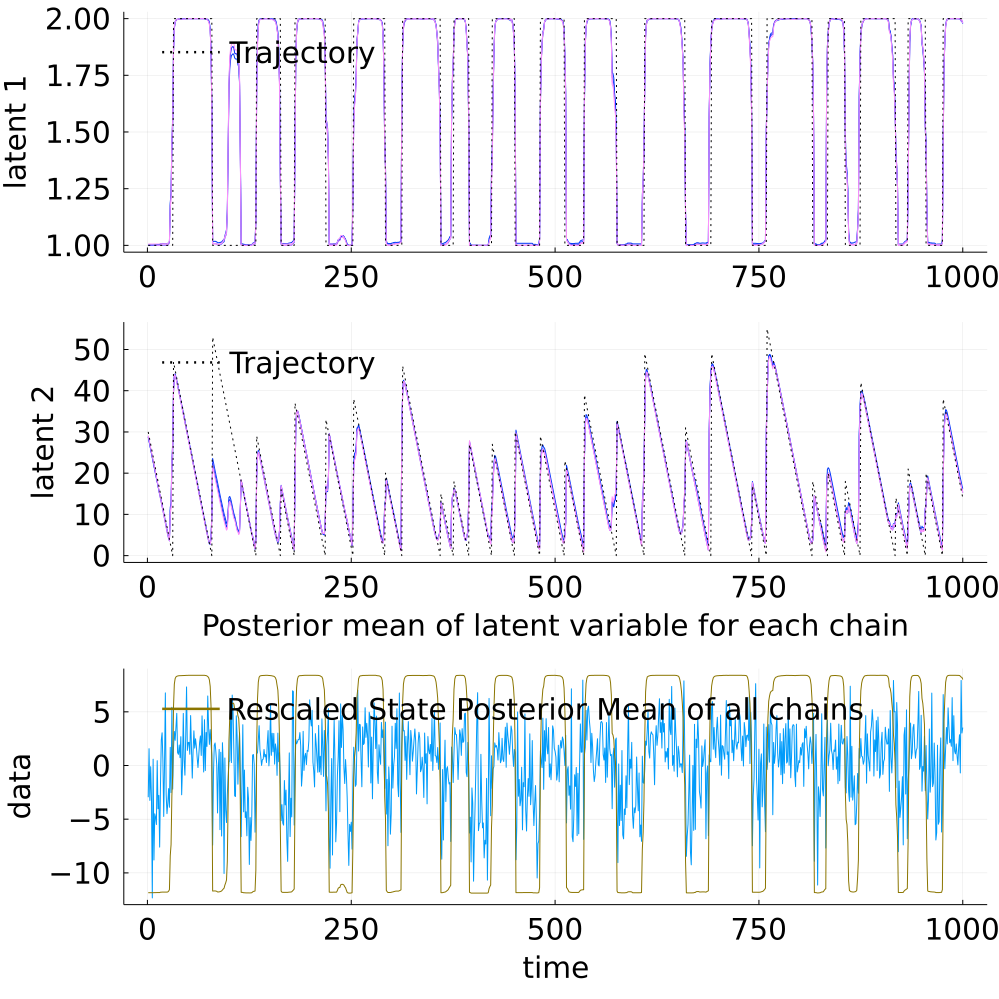}
	\caption{\acrlong{pmcmc} posterior estimates of the filtered latent state trajectory of four chains for the \acrshort{hsmm} in section \ref{sec:hsmm_Experiments}. Parameter used to generate sample data are shown as dashed lines. The bottom plot shows re-scaled posterior means and the observed data.}
	\label{fig:hsmm_Chp5_HSMM_PMCMC_LATENT}
\end{figure}

%%%%%%%%%%%%%%%%%%%%%%%%%%%%%%%%%%%%%%%%%%%%%%%%%%%%%%%%%%%%%%%%%%%%%%%%%
% SMC
\begin{figure}[htp]
	\centering
	\includegraphics[ width=1.\textwidth ]{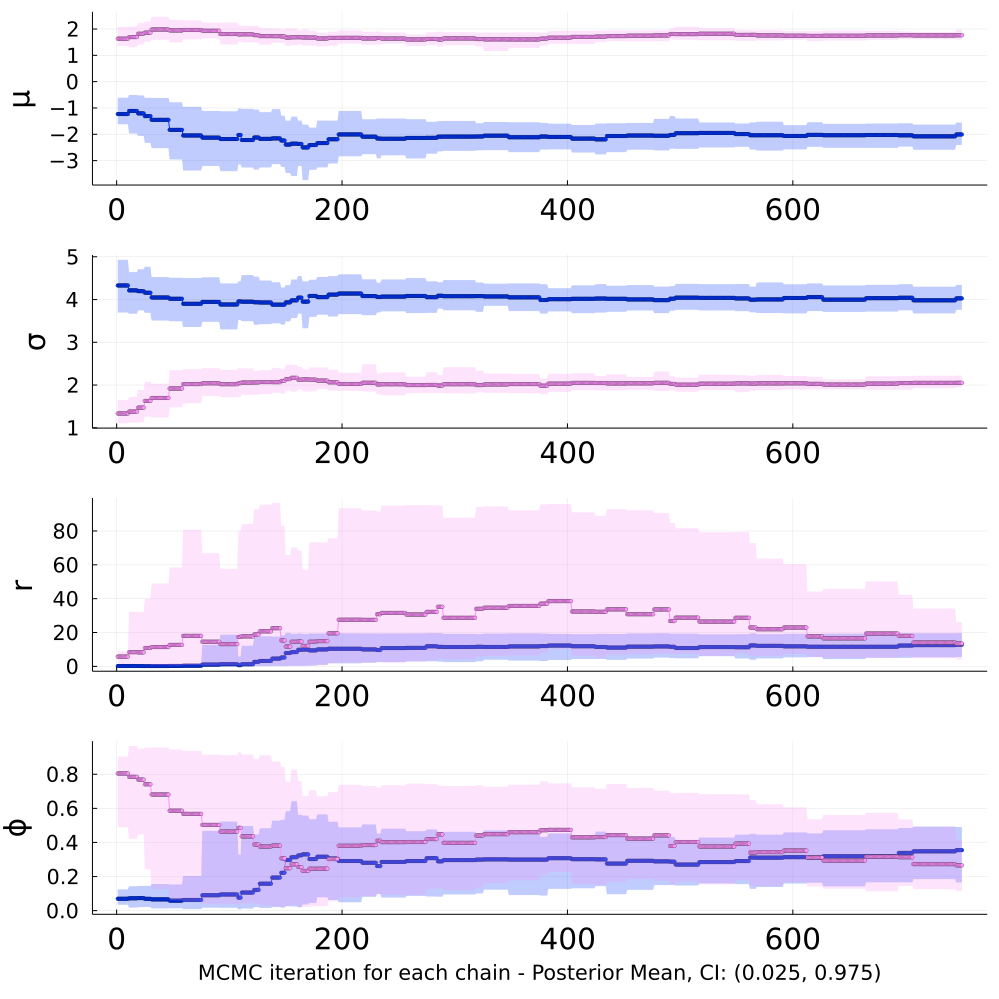}
	\caption{\acrlong{smc2} posterior estimates of the continuous model parameter of $100$ chains for the \acrshort{hsmm} in section \ref{sec:hsmm_Experiments}. The posterior mean and a 95\% \acrlong{ci} are provided for each parameter at each time index. 
    %Parameter used to generate sample data are shown as dashed horizontal lines.
    }
	\label{fig:hsmm_Chp5_HSMM_SMC_CHAIN}
\end{figure}
\begin{figure}[htp]
	\centering
	\includegraphics[ width=1.\textwidth ]{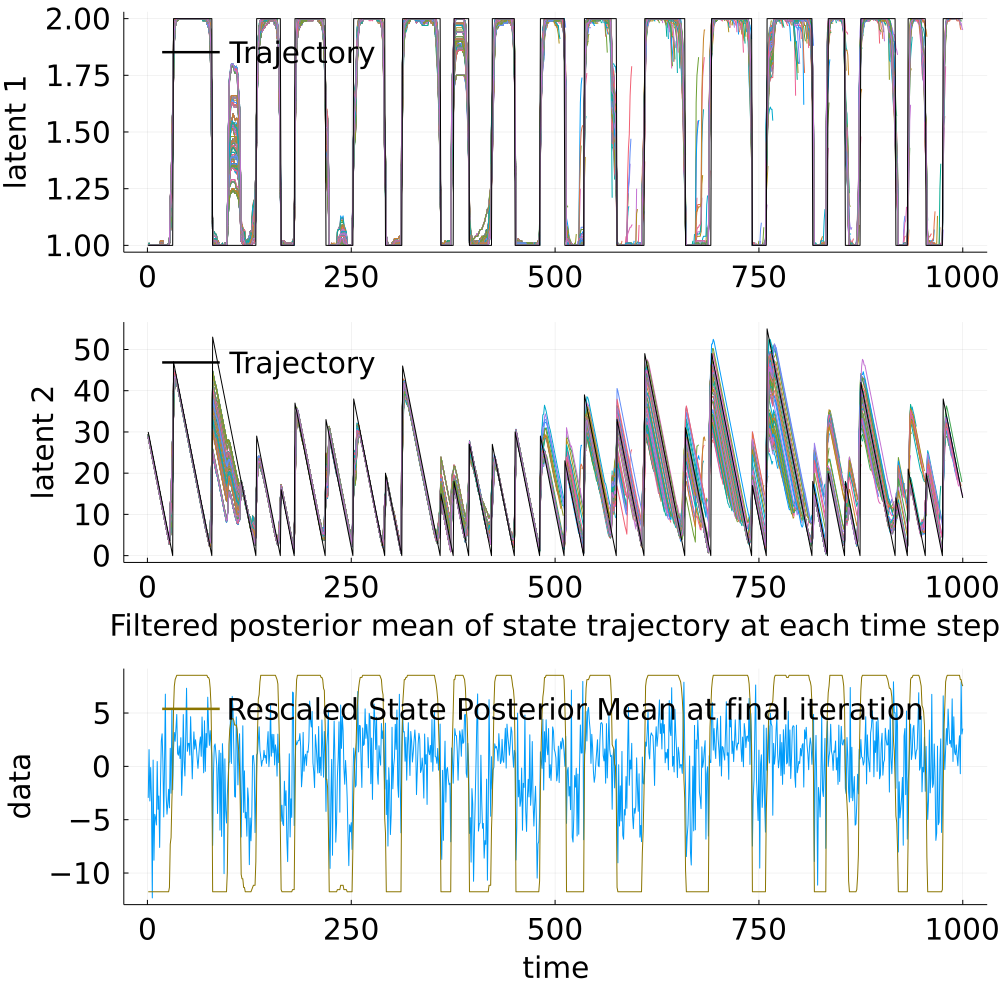}
	\caption{\acrlong{smc} posterior estimates of the filtered latent state trajectory of $100$ chains for the \acrshort{hsmm} in section \ref{sec:hsmm_Experiments} at each time index. 
    The bottom plot shows the underlying observed data and re-scaled posterior mean of the latent state at the final iteration.
    Parameter used to generate sample data are shown as dashed horizontal lines.
 }
	\label{fig:hsmm_Chp5_HSMM_SMC_LATENT}
\end{figure}
\begin{comment}
\begin{figure}[htp]
	\centering
	\includegraphics[ width=1.\textwidth ]{Chp5_HSMM_SMC_PREDICT.png}
	\caption{SMC predictions. \acrlong{smc} posterior predictive samples for the \acrshort{hsmm} in section \ref{sec:hsmm_Experiments}. Black line depicts the realized future value against predictions in gold. The top 2 graph are predictions for the state and duration variables, and the bottom plot shows predictions for the observed data.}
	\label{fig:hsmm_Chp5_HSMM_SMC_PREDICT}
\end{figure}
\end{comment}

% Chp5 - Applications
%%%%%%%%%%%%%%%%%%%%%%%%%%%%%%%%%%%%%%%%%%%%%%%%%%%%%%%%%%%%%%%%%%%%%%%%%
%Real data
\begin{figure}[htp]
	\centering
	\includegraphics[ width=1.\textwidth ]{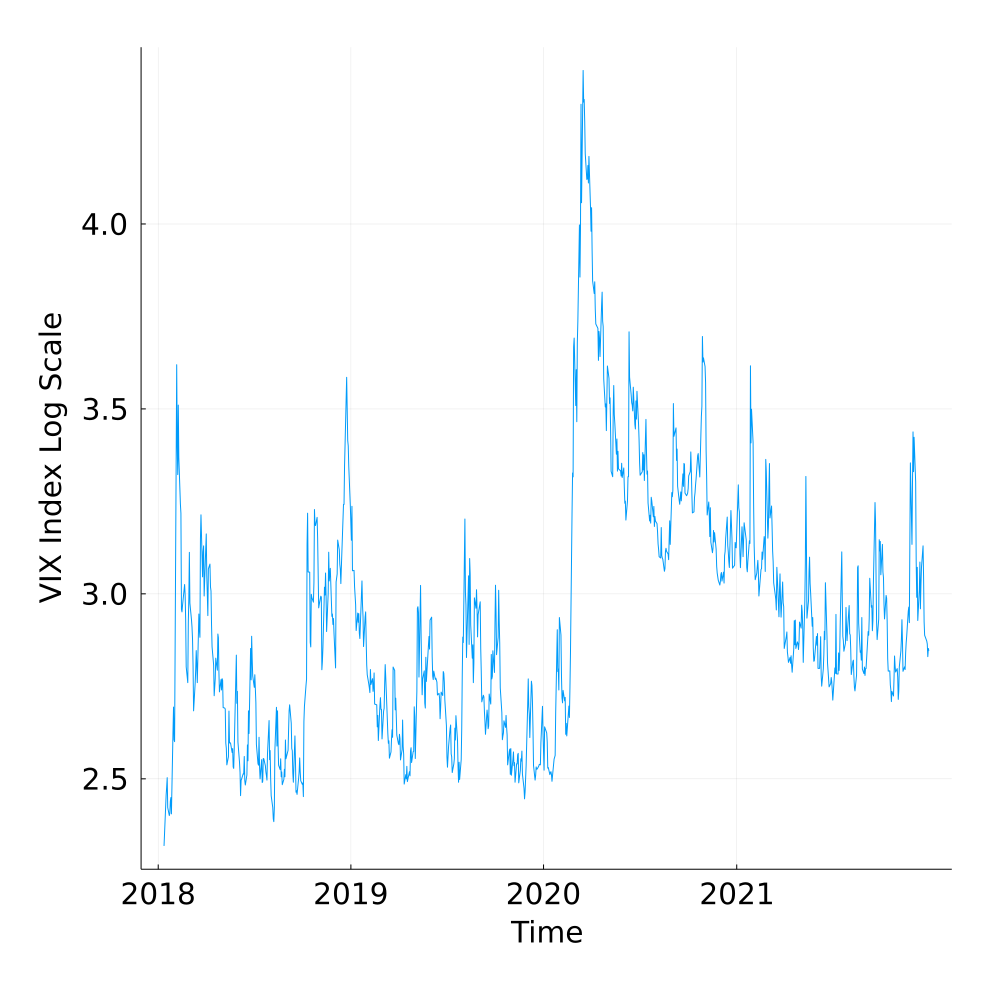}
	\caption{Last 1000 end-of-day data points of the VIX Index used as data in Section \ref{sec:hsmm_Applications}. Data as of January 1st 2022 from the Thomson Reuters database.}
	\label{fig:hsmm_Chp6_realdata}
\end{figure}

% SMC2
\begin{figure}[htp]
	\centering
	\includegraphics[ width=1.\textwidth ]{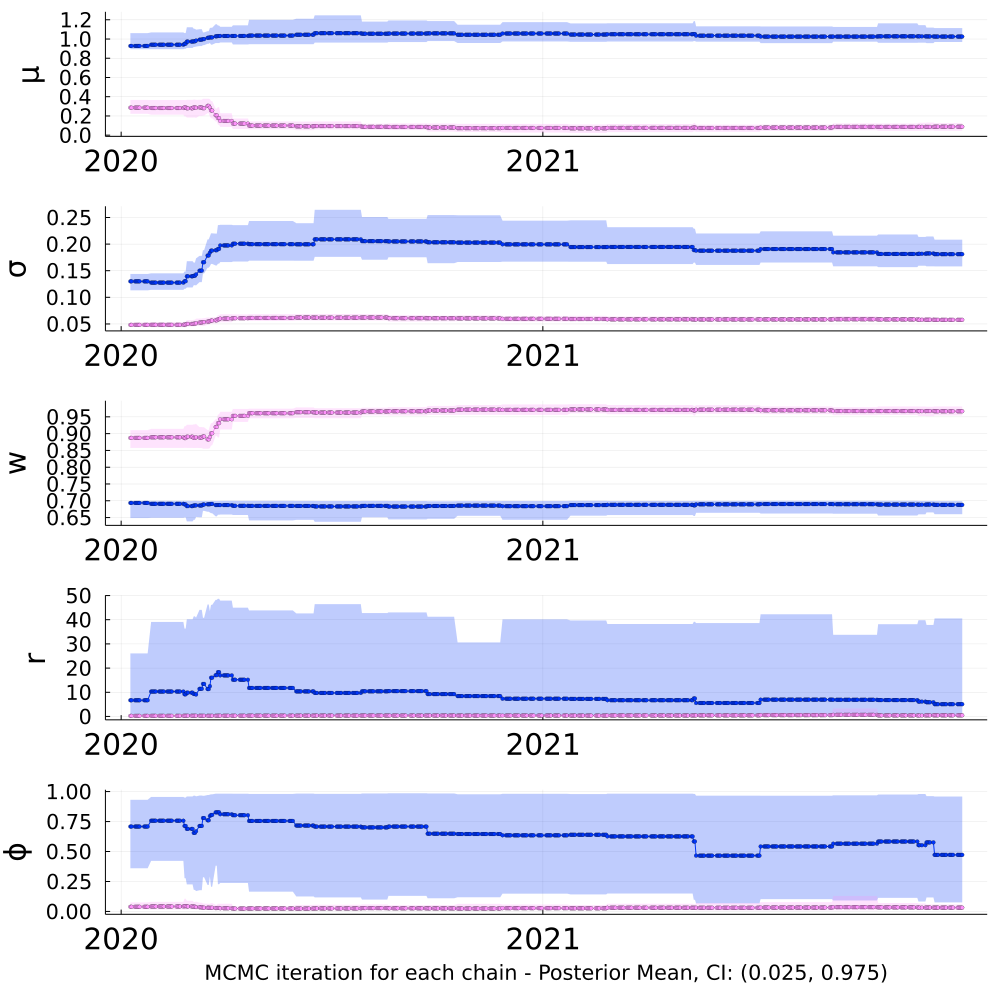}
	\caption{\acrlong{smc} posterior estimates of the continuous model parameter of 100 chains for the AR(1) \acrshort{hsmm} with Negative Binomial duration distribution, discussed in section \ref{sec:hsmm_Applications}. The posterior mean and a 95\% \acrlong{ci} are provided for each parameter at each time index.}
	\label{fig:hsmm_Chp6_SMC2Chain}
\end{figure}
\begin{figure}[htp]
	\centering
	\includegraphics[ width=1.\textwidth ]{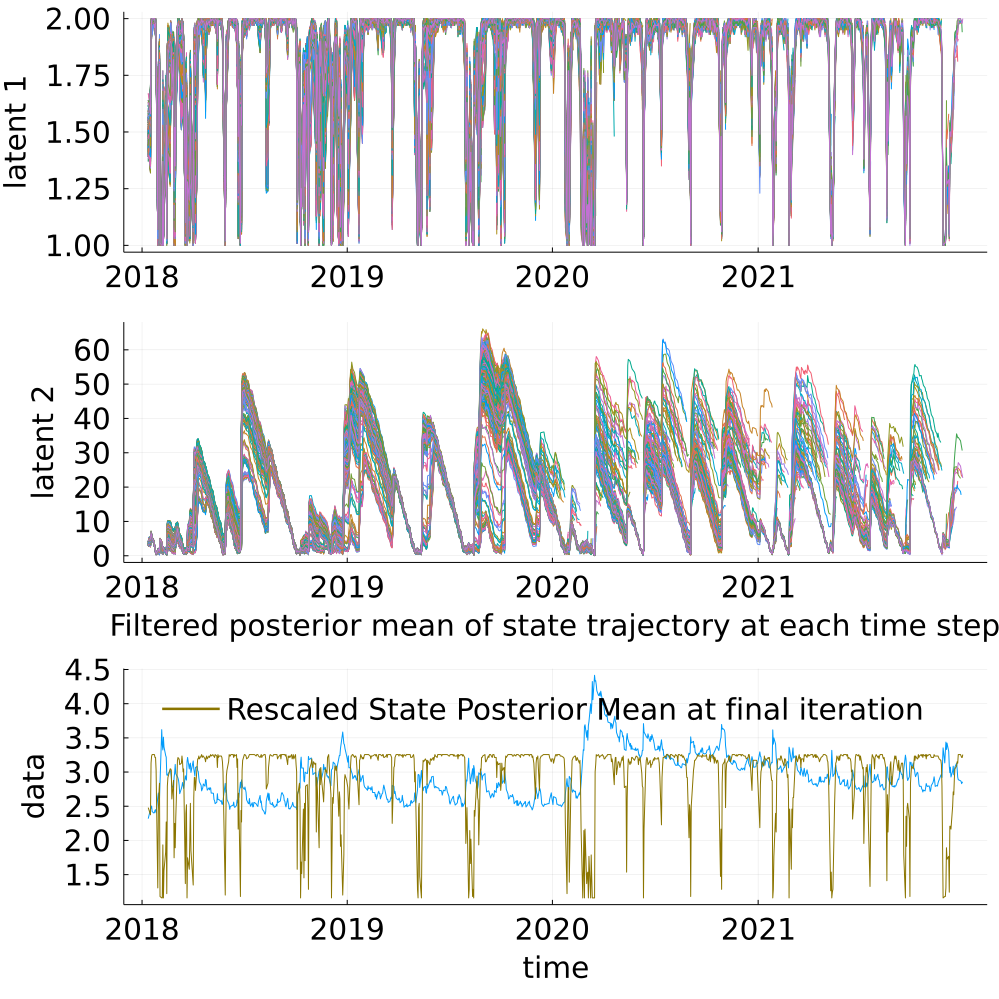}
	\caption{\acrlong{smc} posterior estimates of the filtered latent state trajectory of 100 chains for the AR(1) \acrshort{hsmm} with Negative Binomial duration distribution, discussed in section \ref{sec:hsmm_Applications}. The bottom plot shows the underlying observed data and re-scaled posterior mean of the latent state at the final iteration.}
	\label{fig:hsmm_Chp6_SMC2Latent}
\end{figure}
\begin{figure}[htp]
	\centering
	\includegraphics[ width=1.\textwidth ]{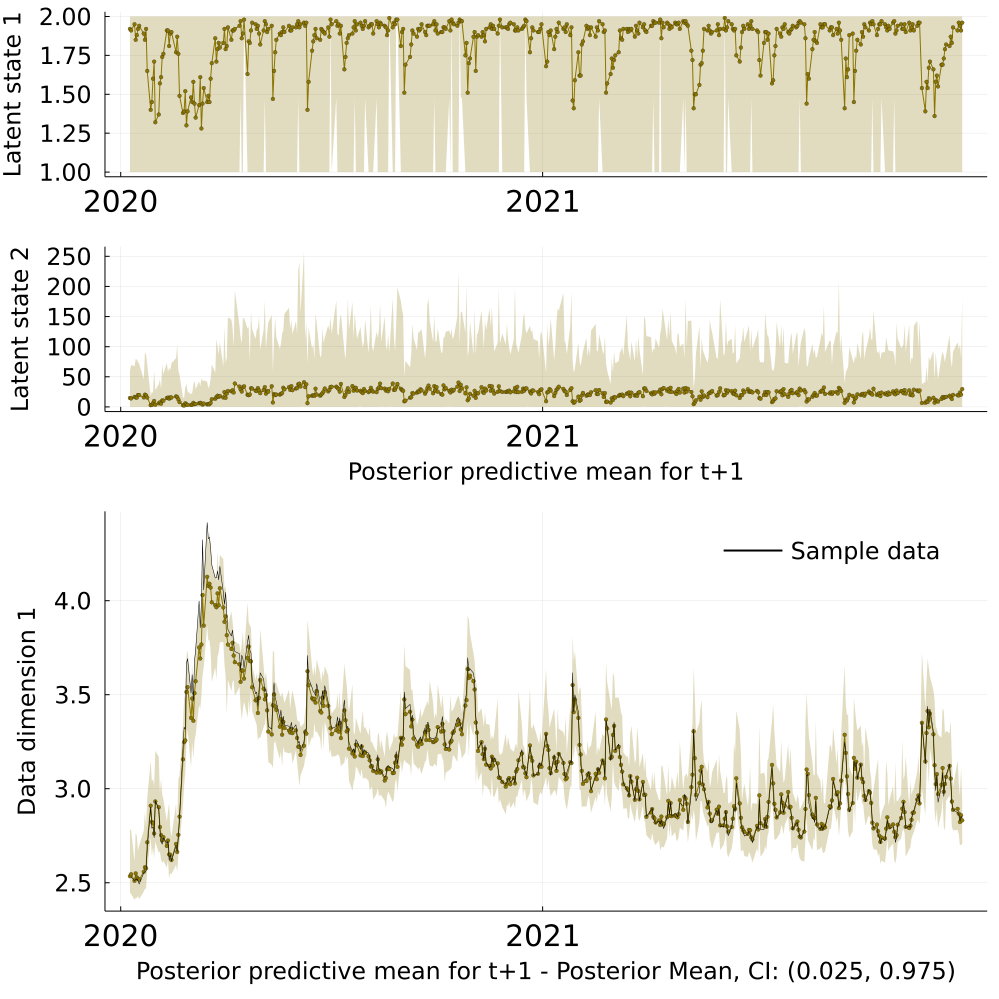}
	\caption{\acrlong{smc2} posterior predictive samples for the AR(1) \acrshort{hsmm} with Negative Binomial duration distribution, discussed in section \ref{sec:hsmm_Applications}. The Black line at the bottom table depicts the realized future value against predictions in gold. The top 2 graph are predictions for the state and duration variables, and the bottom plot shows predictions for the observed data.}
	\label{fig:hsmm_Chp6_SMC2Prediction}
\end{figure}

%%%%%%%%%%%%%%%%%%%%%%%%%%%
%Pseudo Algorithms
\clearpage
%%%%%%%%%%%%%%%%%%%%%%%%%%%%%%%%%%%%%%%%%%%%%%%%%%%%%%%%%%%%%%%%%%%%%%%%%
%%%%%%%%%%%%%%%%%%%%%%%%%%%%%%%%%%%%%%%%%%%%%%%%%%%%%%%%%%%%%%%%%%%%%%%%%
%%%%%%%%%%%%%%%%%%%%%%%%%%%%%%%%%%%%%%%%%%%%%%%%%%%%%%%%%%%%%%%%%%%%%%%%%
\section{Pseudo Algorithms} \label{sec:App_PseudoAlgorithm}

% Chp3 - HSMM
%%%%%%%%%%%%%%%%%%%%%%%%%%%%%%%%%%%%%%%%%%%%%%%%%%%%%%%%%%%%%%%%%%%%%%%%%
% Sample HSMM
\begin{comment}

\begin{algorithm}[H]
	\SetKwData{Left}{left}\SetKwData{This}{this}\SetKwData{Up}{up}
	\SetKwFunction{Union}{Union}\SetKwFunction{FindCompress}{FindCompress}
	\SetKwInOut{Input}{input}\SetKwInOut{Output}{output}
	
	\Input{Parameter $\theta$, number of iterations $T$}
	\Output{Samples for latent states $\{ s_{1:T}, d_{1:T}\}$ and observations $e_{1:T}$}
%	\BlankLine
	\tcp{Initialize:} 
	\emph{$S_1 \sim \pi_\theta(s_1)$, $D_1 \sim \pi_\theta(d_1)$, 
	$E_1 \mid s_{1} \sim g_\theta(e_1 \mid s_{1})$.} \;
	
	\tcp{Forward sample up to T} 
	\For{$t\leftarrow 2$ \KwTo $T$}{
	
    	\begin{equation*} 
        S_t \mid s_{t-1}, d_{t-1} \sim  \begin{cases}
        \delta( s_{t}, s_{t-1}) &\text{ $d_{t-1} > 0$ }\\
        f_\theta( s_t \mid s_{t-1}, d_{t-1} ) &\text{ $d_{t-1} = 0$ }
        \end{cases}
        \end{equation*}
        
        \begin{equation*} 
        D_t \mid s_{t}, d_{t-1} \sim \begin{cases}
        \delta( d_{t}, d_{t-1} - 1) &\text{ $d_{t-1} > 0$ }\\
        h_\theta( s_t \mid s_{t}, d_{t-1}) &\text{ $d_{t-1} = 0$ }
        \end{cases}
        \end{equation*}
        
        \begin{equation*} 
        E_t \mid s_{t} \sim g_\theta(e_t \mid s_{t})
        \end{equation*}
	}	
	\caption{Sampling an EDHMM}
	\label{alg:EDHMM_Sampling}
\end{algorithm}

\end{comment}

% Chp3 - Bayesian Inference
%%%%%%%%%%%%%%%%%%%%%%%%%%%%%%%%%%%%%%%%%%%%%%%%%%%%%%%%%%%%%%%%%%%%%%%%%
% ParticleFilter
\begin{algorithm}[htbp]
	\SetKwData{Left}{left}\SetKwData{This}{this}\SetKwData{Up}{up}
	\SetKwFunction{Union}{Union}\SetKwFunction{FindCompress}{FindCompress}
	\SetKwInOut{Input}{input}\SetKwInOut{Output}{output}\SetKwInOut{Tuning}{tuning parameter}\SetKwInOut{Function}{function} 
	\Input{data $e_{1:T}$, model parameter $\theta$}
	\Output{log-likelihood estimate $\hat{\ell}(\theta) = \log \hat{p}_\theta(e_{1:T})$ and sample $s_{1:T} \sim \hat{p}(s_{1:T} \mid e_{1:T})$ }
	\Tuning{proposal distribution $q$, number of particles $N$}
    \Function{particle filter $pf(e_{1:T}, \theta)$}
    \BlankLine
    
	\tcp{Initialization:}
	\For{$n\leftarrow 1$ \KwTo $N$}{
		Initiate particle \emph{$s^n_{1} \sim \pi_\theta(s_1)$.} 
		
		Compute \emph{$\alpha^n_1(s^n_{1}, e_{1}) = \frac{ p_\theta( e_{1} \mid s^n_{1}) ~ p_\theta( s^n_{1}) }{ q(s^n_1 \mid e_{1}) }$.}
	}
	Normalize weights \emph{$\tilde{\alpha}^i_1 \propto \alpha^i_1(s^n_{1}, e_{1})$ for $i=1:N$, s.t. $\sum_{i=1}^{N} \tilde{\alpha}^i_1 = 1$.} 
	
	Compute log-likelihood increment \emph{$\hat{\ell}(\theta) = \log \frac{1}{N} \sum_{i=1}^{N} \alpha^i_1(s^i_{1}, e_{1})$}

	\tcp{Forward propagation:} 
	\For{$t\leftarrow 2$ \KwTo $T$}{

    \uIf{Resampling required}{
    Sample ancestor $a^n_{t}$ for particle trajectory $s^n_{1:t-1}$ for $n=1$ to $N$ according to normalized weights $\tilde{\alpha}_{t-1}$. 
    }
    \Else{
    Set $a^n_{t} = n$ for $n=1$ to $N$.
    }
    
        \For{$n\leftarrow 1$ \KwTo $N$}{
		    Sample $s^n_{t} \sim q(s_t^n \mid s^{a^n_{t}}_{1:t-1}, e_{1:t})$. 
		    
		    Set $s^n_{1:t} := (s^{a^n_t}_{1:t-1}, s^n_t)$.

		    Calculate incremental weight:
			\begin{equation*} 
				\alpha_t(s^n_{1:t}, e_{1:t}) = \frac{ p_\theta( e_{t} \mid s^n_{1:t}, e_{1:t-1} ) ~p_\theta( s^n_{t} \mid s^n_{1:t-1}, e_{1:t-1} ) }{ q(s^n_t \mid s^n_{1:t-1}, e_{1:t}) }
			%\label{:ParticleWeights}
			\end{equation*}
        }
        Normalize weights \emph{$\tilde{\alpha}^i_t \propto \alpha^i_t(s^i_{1:t}, e_{1:t})$} for $i=1$ to $N$, s.t. $\sum_{i=1}^{N} \tilde{\alpha}^i_t = 1$.
        
        Add incremental weights to log-likelihood: \emph{$\hat{\ell}(\theta) = \hat{\ell}(\theta) + \log \frac{1}{N} \sum_{i=1}^{N} \alpha^i_t(s^i_{1:t}, e_{1:t})$.
		}
	}
	\tcp{Return log-likelihood estimate and particle trajectories:} 
	Draw k with $P(k=i) \propto \tilde{\alpha}^i_T$.
	
	\KwRet $\hat{\ell}(\theta)$ and $s^k_{1:T}$.
	\caption{Standard particle filter }
	\label{alg:ParticleFilter}
\end{algorithm}

%%%%%%%%%%%%%%%%%%%%%%%%%%%%%%%%%%%%%%%%%%%%%%%%%%%%%%%%%%%%%%%%%%%%%%%%% 
% Conditional ParticleFilter
\begin{algorithm}[htbp]
	\SetKwData{Left}{left}\SetKwData{This}{this}\SetKwData{Up}{up}
	\SetKwFunction{Union}{Union}\SetKwFunction{FindCompress}{FindCompress}
	\SetKwInOut{Input}{input}\SetKwInOut{Output}{output}\SetKwInOut{Tuning}{tuning parameter}\SetKwInOut{Function}{function} 
	\Input{Reference $s^{'}_{1:T}$, data $e_{1:T}$, model parameter $\theta$}
	\Output{Log-likelihood estimate $\hat{\ell}(\theta) = \log \hat{p}_\theta(e_{1:T})$ and sample $s_{1:T} \sim \hat{p}(s_{1:T} \mid s^{'}_{t:T}, e_{1:T})$ }
    \Tuning{proposal distribution $q$, number of particles $N$}
    \Function{particle filter $cpf(s^{'}_{1:T}, e_{1:T}, \theta)$}
    \BlankLine
    
	\tcp{Initialization:}
	\For{$n\leftarrow 1$ \KwTo $N-1$}{
		Initiate particle \emph{$s^n_{1} \sim \pi_\theta(s_1)$.} 
		
		Compute \emph{$\alpha^n_1(s^n_{1}, e_{1}) = \frac{ p_\theta( e_{1} \mid s^n_{1}) ~ p_\theta( s^n_{1}) }{ q(s^n_1 \mid e_{1}) }$.}
	}
	Set $s^N_{1} := s^{'}_{1}$ and compute $\alpha^N_1(s^N_{1}, e_{1}) \propto \frac{ p_\theta( e_{1} \mid s^N_{1}) ~ p_\theta( s^N_{1}) }{ q(s^N_1 \mid e_{1}) }$.
	
	Normalize weights \emph{$\tilde{\alpha}^i_1 \propto \alpha^i_1(s^n_{1}, e_{1})$ for $i=1:N$, s.t. $\sum_{i=1}^{N} \tilde{\alpha}^i_1 = 1$.} 
	
	Compute log-likelihood increment \emph{$\hat{\ell}(\theta) = \log \frac{1}{N} \sum_{i=1}^{N} \alpha^i_1(s^i_{1}, e_{1})$}

	\tcp{Forward propagation:} 
	\For{$t\leftarrow 2$ \KwTo $T$}{
	
	\uIf{Resampling required}{
    Sample ancestor $a^n_{t}$ for particle trajectory $s^n_{1:t-1}$ for $n=1$ to $N-1$ according to normalized weights $\tilde{\alpha}_{t-1}$. 
    
    Set $a^N_{t} = k$, with $P(k=i) \sim \alpha^i_{t-1}(s^i_{1:t-1}, e_{1:t-1}) ~p_\theta(s^{'}_{t:T}, e_{t:T} \mid s^{i}_{1:t-1}, e_{1:t-1})$.
    }
    \Else{
    Set $a^n_{t} = n$ for $n=1$ to $N$.
    }
  
        \For{$n\leftarrow 1$ \KwTo $N$}{
		    Sample $s^n_{t} \sim q(s_t^n \mid s^{a^n_{t}}_{1:t-1}, e_{1:t})$. 
		    
		    Set $s^n_{1:t} := (s^{a^n_t}_{1:t-1}, s^n_t)$.

		    Calculate incremental weight:
			\begin{equation*} 
				\alpha_t(s^n_{1:t}, e_{1:t}) = \frac{ p_\theta( e_{t} \mid s^n_{1:t}, e_{1:t-1} ) ~p_\theta( s^n_{t} \mid s^n_{1:t-1}, e_{1:t-1} ) }{ q(s^n_t \mid s^n_{1:t-1}, e_{1:t}) }
			%\label{:ParticleWeights}
			\end{equation*}
        }
        Normalize weights \emph{$\tilde{\alpha}^i_t \propto \alpha^i_t(s^i_{1:t}, e_{1:t})$} for $i=1$ to $N$, s.t. $\sum_{i=1}^{N} \tilde{\alpha}^i_t = 1$.
        
        Add incremental weights to log-likelihood: \emph{$\hat{\ell}(\theta) = \hat{\ell}(\theta) + \log \frac{1}{N} \sum_{i=1}^{N} \alpha^i_t(s^i_{1:t}, e_{1:t})$.
		}
	}
	\tcp{Return log-likelihood estimate and particle trajectories:} 
	Draw k with $P(k=i) \propto \tilde{\alpha}^i_T$.
	
	\KwRet $\hat{\ell}(\theta)$ and $s^k_{1:T}$.
	\caption{Conditional particle filter with ancestor sampling}
	\label{alg:ConditionalParticleFilter}
\end{algorithm}

%%%%%%%%%%%%%%%%%%%%%%%%%%%%%%%%%%%%%%%%%%%%%%%%%%%%%%%%%%%%%%%%%%%%%%%%%
% Metropolis Hastings
\begin{algorithm}[htbp]
	\SetKwData{Left}{left}\SetKwData{This}{this}\SetKwData{Up}{up}
	\SetKwFunction{Union}{Union}\SetKwFunction{FindCompress}{FindCompress}
	\SetKwInOut{Input}{input}\SetKwInOut{Output}{output}\SetKwInOut{Tuning}{tuning parameter}\SetKwInOut{Function}{function} 
	\Input{data $e_{1:T}$, current model parameter $\theta$}
	\Output{model parameter $\theta \sim p(\theta \mid e_{1:T})$} %, $(\theta^i)_{i = 1:N}$. }
    \Tuning{proposal distribution $f$}
    \Function{MCMC Kernel $K_{mh}(e_{1:T}, \theta)$}
    \BlankLine
    
%	\tcp{Initialization:}
%	Initiate parameter vector \emph{$\theta \sim p(\theta)$.} 
    
%	\tcp{Transition:} 
%	\For{$n\leftarrow 1$ \KwTo $N$}{
        Propose \emph{$\theta^{\star} \sim f(\theta^{\star} \mid \theta )$}.
        
        Set $\theta := \theta^{\star}$ with acceptance probability $min(1, a( \theta^{\star}, \theta ))$, where
	    \begin{equation*}
	    \begin{split}
            a( \theta^{\star}, \theta ) &= \frac{ p(\theta \mid e_{1:T}) ~ f( \theta  \mid \theta^{\star})}{p(\theta \mid e_{1:T}) ~ f( \theta^{\star} \mid \theta )} \\
            &= \frac{p_{\theta^{\star}}(e_{1:T} ) ~ p(\theta^{\star}) ~ f( \theta  \mid \theta^{\star})}{p_{\theta }(e_{1:T} ) ~ p(\theta ) ~ f( \theta^{\star} \mid \theta )}.
	    \end{split} 
        \end{equation*}
        
        \KwRet $\theta$.
%	}
	\caption{Metropolis Hastings (MH) Kernel}
	\label{alg:MH}
\end{algorithm}

%%%%%%%%%%%%%%%%%%%%%%%%%%%%%%%%%%%%%%%%%%%%%%%%%%%%%%%%%%%%%%%%%%%%%%%%%
% Hamiltonian Monte Carlo
\begin{algorithm}[htbp]
	\SetKwData{Left}{left}\SetKwData{This}{this}\SetKwData{Up}{up}
	\SetKwFunction{Union}{Union}\SetKwFunction{FindCompress}{FindCompress}
	\SetKwInOut{Input}{input}\SetKwInOut{Output}{output}\SetKwInOut{Tuning}{tuning parameter}\SetKwInOut{Function}{function} 
	\Input{data $e_{1:T}$, current model parameter $\theta$}
	\Output{model parameter $\theta \sim p(\theta \mid e_{1:T})$} %, $(\theta^i)_{i = 1:N}$. }
    \Tuning{Mass matrix $M$, stepsize $\epsilon$, number of leapfrog steps $L$.}
    \Function{MCMC Kernel $K_{HMC}(e_{1:T}, \theta)$}
    \BlankLine

    Sample $\rho \sim MvNormal(0, M)$ and set $(\theta^{\star}, \rho^{\star}) := (\theta, \rho)$.
        
    \For{$ i\leftarrow 1$ \KwTo $L$}{
        $\theta^{\star}, \rho^{\star} = Leapfrog(\theta^{\star}, \rho^{\star}, M, \epsilon)$
    }    
        Set $\theta := \theta^{\star}$ with acceptance probability $min(1, a( \theta^{\star}, \theta ))$, where
	    \begin{equation*}
%	    \begin{split}
            a( \theta^{\star}, \theta ) = exp( H(\rho, \theta) - H(\rho^{\star}, \theta^{\star}) ) %\\
            %&= exp( (-log ~p(\rho \mid \theta)  - log ~p_{\theta}(e_{1:T}) ) - (-log ~p(\rho^{\star} \mid \theta^{\star})  - log ~p_{\theta^{\star}}(e_{1:T}) ) ).
%	    \end{split} 
        \end{equation*}
        
        \KwRet $\theta$.
    
    \BlankLine
    \SetKwFunction{FMain}{Leapfrog}
    \SetKwProg{Fn}{Function}{:}{}
    \Fn{\FMain{$\theta_{t}, \rho_{t}, M, \epsilon$}}{
        $\rho_{t + \frac{\epsilon}{2} } \leftarrow \rho_{t} + \frac{\epsilon}{2}\frac{\partial log~p(\theta \mid e_{1:T}) }{\partial \theta}(\theta_{t})$

        $\theta_{t+\epsilon} \leftarrow \theta_{t} + \epsilon M^{-1} \rho_{t + \frac{\epsilon}{2} }$

        $\rho_{t+\epsilon} \leftarrow \rho_{t + \frac{\epsilon}{2} } + \frac{\epsilon}{2}\frac{\partial log~p(\theta \mid e_{1:T}) }{\partial \theta}(\theta_{t+\epsilon})$
        
        \KwRet $\theta_{t+\epsilon}, \rho_{t+\epsilon}$.
    }

	\caption{Hamiltonian Monte Carlo (HMC) Kernel}
	\label{alg:HMC}
\end{algorithm}

%%%%%%%%%%%%%%%%%%%%%%%%%%%%%%%%%%%%%%%%%%%%%%%%%%%%%%%%%%%%%%%%%%%%%%%%%
% Particle MCMC
\begin{algorithm}[htbp]
	\SetKwData{Left}{left}\SetKwData{This}{this}\SetKwData{Up}{up}
	\SetKwFunction{Union}{Union}\SetKwFunction{FindCompress}{FindCompress}
	\SetKwInOut{Input}{input}\SetKwInOut{Output}{output}\SetKwInOut{Tuning}{tuning parameter}\SetKwInOut{Function}{function} 
	\Input{current state trajectory $s_{1:T}$, data $e_{1:T}$, current model parameter $\theta$}
	\Output{model parameter $\theta$ and state trajectory $s_{1:T}$, $(\theta, s_{1:T}) \sim p(\theta, s_{1:T} \mid e_{1:T})$ }
    \Tuning{MCMC kernel $K_{mcmc}$, particle filter $pf$, number of iterations $N$}
    \Function{PMCMC kernel $K_{pmh}(e_{1:T}, s_{1:T}, \theta)$}
    \BlankLine
    
%	\tcp{Initialization:}
%        Initiate parameter vector \emph{$\theta \sim p(\theta)$.}
		
%		Run particle filter to obtain $\hat{p}_\theta(e_{1:T})$ and $s_{1:T} \sim \hat{p}_\theta(s_{1:T} \mid e_{1:T})$.

%	\tcp{Transition:} 
%	\For{$n\leftarrow 1$ \KwTo $N$}{
    Propose \emph{$\theta^{\star} \sim K_{mcmc}(e_{1:T}, \theta )$}
     
	Run particle filter $pf$ to obtain $\hat{p}_{\theta^{\star}}(e_{1:T})$ and $s^{\star}_{1:T} \sim \hat{p}_{\theta^{\star}}(s_{1:T}^{\star} \mid e_{1:T})$.
	
	Set $(\theta, s_{1:T}) := (\theta^{\star}, s^{\star}_{1:T})$ with acceptance probability $min(1, a( \theta^{\star}, \theta ))$, where
	\begin{equation}
        a( \theta^{\star}, \theta ) = \frac{\hat{p}_{\theta^{\star}}(e_{1:T} ) ~ p(\theta^{\star}) ~ f_{mcmc}( \theta  \mid \theta^{\star})}{\hat{p}_{\theta }(e_{1:T} ) ~ p(\theta ) ~ f_{mcmc}( \theta^{\star} \mid \theta )}
    \end{equation}

    \KwRet $(\theta, s_{1:T})$
%	}

	\caption{Particle Metropolis Hastings Kernel}
	\label{alg:PMCMC}
\end{algorithm}

%%%%%%%%%%%%%%%%%%%%%%%%%%%%%%%%%%%%%%%%%%%%%%%%%%%%%%%%%%%%%%%%%%%%%%%%%
% Particle Gibbs
\begin{algorithm}[htbp]
	\SetKwData{Left}{left}\SetKwData{This}{this}\SetKwData{Up}{up}
	\SetKwFunction{Union}{Union}\SetKwFunction{FindCompress}{FindCompress}
	\SetKwInOut{Input}{input}\SetKwInOut{Output}{output}\SetKwInOut{Tuning}{tuning parameter}\SetKwInOut{Function}{function} 
	\Input{reference trajectory $s_{1:T}$, data $e_{1:T}$, current model parameter $\theta$}
	\Output{model parameter $\theta$ and state trajectory $s_{1:T}$, $(\theta, s_{1:T}) \sim p(\theta, s_{1:T} \mid e_{1:T})$ }
    \Tuning{conditional particle filter $cpf$}
    \Function{PMCMC kernel $K_{pgibbs}(e_{1:T}, s_{1:T}, \theta)$}
    \BlankLine

%	\tcp{Initialization:}
%        Initiate parameter vector \emph{$\theta \sim p(\theta)$.}
		
%		Run particle filter to obtain $s_{1:T} \sim \hat{p}_\theta(s_{1:T} \mid e_{1:T})$.

%	\tcp{Transition:} 
%	\For{$n\leftarrow 1$ \KwTo $N$}{
    Propose \emph{$\theta^{\star} \sim p_\theta(\theta^{\star} \mid s_{1:T}, e_{1:T})$} %\sim f(\theta^{\star} \mid \theta )$}
     
	Run a conditional particle filter $cpf$ to obtain $s^{\star}_{1:T} \sim \hat{p}_{\theta^{\star}}(s_{1:T}^{\star} \mid s_{1:T}, e_{1:T})$. 
	
	Set $(\theta, s_{1:T}) := (\theta^{\star}, s^{\star}_{1:T})$ 
%	}
    
    \KwRet $(\theta, s_{1:T})$
	\caption{Particle Gibbs Kernel}
	\label{alg:PGibbs}
\end{algorithm}

%%%%%%%%%%%%%%%%%%%%%%%%%%%%%%%%%%%%%%%%%%%%%%%%%%%%%%%%%%%%%%%%%%%%%%%%%
% SMC2
\begin{algorithm}[htbp]
	\SetKwData{Left}{left}\SetKwData{This}{this}\SetKwData{Up}{up}
	\SetKwFunction{Union}{Union}\SetKwFunction{FindCompress}{FindCompress}
	\SetKwInOut{Input}{input}\SetKwInOut{Output}{output}\SetKwInOut{Tuning}{tuning parameter}\SetKwInOut{Function}{function} 
	\Input{Data $e_{1:T}$}
	\Output{model parameter $\theta$ and state trajectory $s_{1:t}$, $(\theta^i,s^i_{1:t})_{i = 1:N} \sim p(\theta^i, s_{1:t}^i \mid e_{1:t})$ for $t = 1$ to $T$  }
	\Tuning{number of particles $N$, N particle filter $(pf_i)_{i=1:N}$, N PMCMC kernel $(K_{pmcmc, i})_{i=1:N}$}
    \Function{SMC2 sampler $smc^2(e_{1:T})$}
    \BlankLine
    
	\tcp{Initialization:}
        \For{$n\leftarrow 1$ \KwTo $N$}{

            Initiate parameter vector \emph{$\theta_n \sim p(\theta_n)$.}
		
		    Run particle filter $pf_n$ to obtain $s^n_{1:t_0} \sim \hat{p}_{\theta_n}(s^n_{1:t_0} \mid e_{1:t_0})$.
		    
	    }

	\tcp{Transition:} 
	\For{$t\leftarrow t_0+1$ \KwTo $T$}{
	
        \uIf{Resampled at t-1}{
    
        \For{$n\leftarrow 1$ \KwTo $N$}{
    
            Run particle filter $pf_n$ to obtain $\hat{p}_{\theta^n}(e_{1:t} \mid e_{1:t-1})$ and $s^{n}_{1:t} \sim \hat{p}_{\theta^{n}}(s^{n}_{1:t} \mid e_{1:t})$.
    
        }
    }
    \Else{
        \For{$n\leftarrow 1$ \KwTo $N$}{
        
            Propagate particle filter $pf_n$ forward to obtain $\hat{p}_{\theta^n}(e_{1:t} \mid e_{1:t-1})$ and $s^n_{t} \sim \hat{p}_{\theta^n}( s^n_{t} \mid s^n_{1:t-1}, e_{1:t})$.

            Set $s^n_{1:t} := (s^{n}_{1:t-1}, s^n_t)$.
        }
    }
    Normalize incremental weights \emph{$\tilde{\alpha}^n_t \propto \hat{p}_{\theta^n}(e_{1:t} \mid e_{1:t-1})$} for $n=1$ to $N$, s.t. $\sum_{n=1}^{N} \tilde{\alpha}^n_t = 1$.\\ 
    Compute an estimate for the incremental marginal likelihood:
    \begin{equation}
    \hat{L}_t = \hat{p}(e_t \mid e_{1:t-1}) = \sum_{n=1}^N \tilde{\alpha}^n_t \hat{p}_{\theta^n}(e_{1:t} \mid e_{1:t-1}).   
    \label{eq:App_IncrementalMarginalLikelihood}
    \end{equation}

    \uIf{Resampling required}{
    
        \For{$n\leftarrow 1$ \KwTo $N$}{
            Draw k with $P(k=i) \propto \tilde{\alpha}^i_t$.
    
            Propose $(\theta^{\star}, s^{\star}_{1:t}) \sim K_{pmcmc, k}(e_{1:t}, s^k_{1:t}, \theta^k)$.
    
            Set $(\theta^n, s^n_{1:t}) := (\theta^{\star}, s^{\star}_{1:t})$
    
        }
    }
	
	}
	
    \KwRet $(\theta^i,s^i_{1:t})_{i = 1:N}$ for $t = 1$ to $T$.
    
	\caption{Sequential Monte Carlo Squared algorithm}
	\label{alg:SMC2}
\end{algorithm}

%%%%%%%%%%%%%%%%%%%%%%%%%%%

\vspace{\fill}

%\vfill\eject
\end{document}